\documentclass[twocolumn,groupedaddress,amsmath,aps,longbibliography,nofootinbib]{revtex4-1}
\usepackage{hyperref}
\usepackage{graphicx}
\usepackage{amsmath,amsfonts}
\usepackage{dcolumn}
\usepackage{bm}
\usepackage{color}
\usepackage{multirow}
\usepackage{float}
\usepackage{url}
\usepackage{physics}
\usepackage{subfigure}
\usepackage[utf8x]{inputenc}
\usepackage{soul}
\usepackage[normalem]{ulem}
\usepackage{soul}
\begin{document}

\title{Electronic structure, electron-phonon coupling and superconductivity in noncentrosymmetric ThCoC$_2$ from {\it ab initio} calculations}
\author{Gabriel Kuderowicz}
\author{Pawe\l{} W\'ojcik}
\author{Bartlomiej Wiendlocha}
\email{wiendlocha@fis.agh.edu.pl}
\affiliation{Faculty of Physics and Applied Computer Science, AGH University of Science and Technology, Aleja Mickiewicza 30, 30-059 Krakow, Poland}

\date{\today}

\begin{abstract}
Superconductors without inversion symmetry in their crystal structure are known to exhibit unconventional properties. 
Recently, based on the 
measured temperature dependence of the magnetic field penetration depth, 
superconductivity in noncentrosymmetric $\mathrm{ThCoC_2}$ was proposed to be a nodal $d$-wave and mediated by the spin fluctuations.
Moreover, a non-BCS behavior of the temperature dependence of the electronic specific heat and the magnetic upper critical field were reported. 
In this work, the electronic structure, phonons and 
electron-phonon coupling are studied in $\mathrm{ThCoC_2}$ on the basis of \textit{ab initio} computations.
The effect of the spin-orbit coupling on the electronic structure and electron-phonon interaction is analyzed, and a large splitting of the electronic band structure is found.
The calculated electron-phonon coupling constant $\lambda = 0.59$ 
remains in decent agreement with the experimental estimates, {suggesting that the electron-phonon interaction is strong enough to explain superconductivity with $T_c \simeq 2.5$~K.
Nevertheless, we show that the conventional isotropic Eliashberg formalism is unable to describe the thermodynamic properties of the superconducting state, as 
calculated temperature dependence of the electronic specific heat and magnetic penetration depth deviate from experiments, which is likely driven by 
the strong spin-orbit coupling and inversion symmetry breaking.
In addition, to shed more light on the pairing mechanism, we propose to measure the carbon isotope effect, as our calculations based on the electron-phonon coupling 
predict the observation of the isotope effect with an exponent $\alpha \simeq 0.15$.}
\end{abstract}

\maketitle

\section{Introduction}

In recent years, noncentrosymmetric superconductors (NCSs) have attracted a growing interest as host materials for studying the interplay between the 
spin-orbit coupling (SOC) and the superconducting state of matter. The lack of inversion symmetry in NCSs together with the corresponding
antisymmetric spin-orbit coupling (ASOC) removes the parity constraint on the Cooper pair and allows for a mixture of spin-singlet and spin-triplet 
states~\cite{mixedstate1,mixedstate2,mixedstate3,mixedstate4,mixedstate5}. This 
unusual Cooper pair formation results in an unconventional superconductivity which may manifest itself by the gap with nodal lines or points, 
multiband effects, or the extraordinary high magnetic critical field. 
Among NCSs discovered up-to-date, a special place is 
given for the 
heavy fermion superconductors with the first discovered $\mathrm{CePt_3Si}$~\cite{Bauer2004} as well as $\mathrm{CeRhSi_3}$~\cite{Kimura2007} or 
$\mathrm{CeCoGe_3}$~\cite{Settai2008}. 
The superconducting state in these compounds coexists with the magnetic ordering and is generated by the strong 
electronic correlations which make the physical effects coming from the solely noncentrosymmetric structure difficult to extract. 
To overcome this obstacle, 
most recent studies are directed towards weakly correlated NCSs, including $\mathrm{Li_2(Pd,Pt)_3B}$~\cite{Togano2004,Badica2005,Nishiyama2007,Yuan2006,Lee2005}, $\mathrm{Mg_{10}Ir_{19}B_{16}}$~\cite{Klimczuk2007}, $\mathrm{La(Ir,Rh)(P,As)}$~\cite{Qi2014}, 
$\mathrm{Ru_7B_3}$~\cite{Fang2009,Kase2009},
$\mathrm{Y_2C_3}$~\cite{Chen2011,Kuroiwa2008}, $\mathrm{Mo_3Al_2C}$~\cite{Bauer2010},
$\mathrm{(Nb,Ta)Rh_2B_2}$~\cite{tarh2b2}, $\mathrm{(Nb,Ta)Ir_2B_2}$~\cite{nbir2b2},
or
$\mathrm{LaNiC_2}$~\cite{Lee1996,Pecharsky1998,Iwamoto1998,Quintanilla2010,Bonalde2011,Chen2013A,Katano2014,lanic2,Landaeta2017,Annett2018},  which is isostructural and isoelectronic with $\mathrm{ThCoC_2}$.

\begin{figure}[b]
	\centering
	\includegraphics[width=\columnwidth]{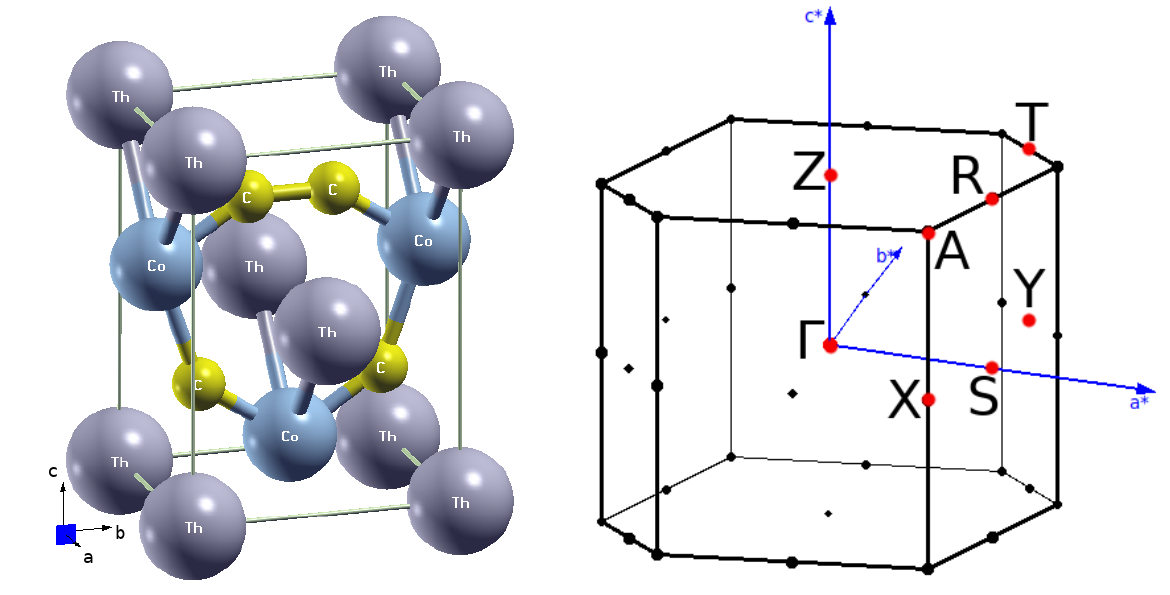}
	\caption{The unit cell of $\mathrm{ThCoC_2}$ visualized with XCrysDen \cite{Kokalj2003} and the Brilloiun zone (BZ) with high symmetry 
points.\label{plot_struct}}	
\end{figure}

Although $\mathrm{LaNiC_2}$ belongs to a group of superconductors with relatively weak electron-phonon coupling ($T_c = 2.8$~K, $\lambda \simeq 0.50$), 
it exhibits non-BCS 
superconducting properties, reported independently in the specific heat~\cite{Lee1996} and penetration depth measurements~\cite{Quintanilla2010,Bonalde2011,Chen2013A}. Despite this, up 
to now there is no general agreement whether the unconventional phase in $\mathrm{LaNiC_2}$ is caused by the multiband Fermi surface, 
spin-triplet pairing or time-reversal symmetry breaking \cite{Landaeta2017,Annett2018}. 
 The most recent muon spin rotation ($\mu$SR) measurements on single crystals of LaNiC$_2$ point to the two nodeless gaps with the broken time-reversal symmetry ~\cite{lanic2-2gaps}.
Another striking behavior of $\mathrm{LaNiC_2}$ has been 
discovered under the external pressure, which initially increases the critical temperature to suppress superconductivity above 7 GPa~\cite{Katano2014}. 
The evolution of $T_c$ with pressure and the 
thermodynamic properties of $\mathrm{LaNiC_2}$, with the special attention paid to their non-BCS character, have been analyzed in Ref.~\cite{lanic2}, on the basis of the {\it ab initio} calculations and the isotropic Eliashberg formalism. 
It was found that the measured non-BCS temperature dependence of the London magnetic penetration depth~\cite{Chen2013A} is close to that predicted by the Eliashberg theory, 
which made it more difficult to clearly classify $\mathrm{LaNiC_2}$ as an unconventional superconductor.

Recent years directed studies on NCSs towards the $\mathrm{ThCoC_2}$ compound, which crystallizes in the same base centered orthorhombic structure (\textit{Amm2}, spacegroup no. 38), as $\mathrm{LaNiC_2}$. Unit cell is shown in Fig.~\ref{plot_struct}.
They both belong to a large series of rare-earth carbides $R$CoC$_2$ and $R$NiC$_2$, recently intensively studied mainly due to their interesting magnetic properties \cite{Kolincio2017,Roman2018,Steiner2018,Roman2018a,Kolincio2019}. 
$\mathrm{ThCoC_2}$ is a nonmagnetic, type-II superconductor with a critical temperature $T_c \simeq 2.5$~K (reported $T_c$ slightly varies between the samples, being $2.65$~K in Ref.~\cite{Grant2014}, $2.55$~K in Ref.~\cite{Grant2017} and $2.3$~K in Ref.~\cite{Bhattacharyya2019}).
It shows several non-BCS thermodynamic properties in the superconducting state.
The upper magnetic critical field ($H_{c2}$) measured as a function of temperature 
exhibits a positive curvature~\cite{Grant2014} which is commonly attributed to the interband coupling occurring in multiband systems, similar to $\mathrm{MgB_2}$~\cite{Takano2001,Shulga1998,Chen2013A}.
The electronic specific heat measurement reveals a strong non-BCS behavior with the normalized specific heat jump $\Delta C_e /\gamma T_c=0.86$ and a significant 
deviation from the exponential trend of $C_e(T)$ at low temperatures~\cite{Grant2014}. 
Importantly, the normalized residual Sommerfeld coefficient $\gamma_0$ exhibits a square 
root magnetic field dependence $\gamma_0 \sim \sqrt{H}$~\cite{Grant2017}, which is commonly considered a hallmark of nodes in the superconducting order parameter. 
The nodal line scenario has been recently strengthen by the $\mu$SR measurement where the temperature dependence of the magnetic penetration depth was well fitted assuming a $d-$wave superconducting gap symmetry~\cite{Bhattacharyya2019}. Moreover, spin-fluctuation mechanism of electron paring was proposed in that work.

\begin{figure*}[t]
	\centering
	\includegraphics[width=\textwidth]{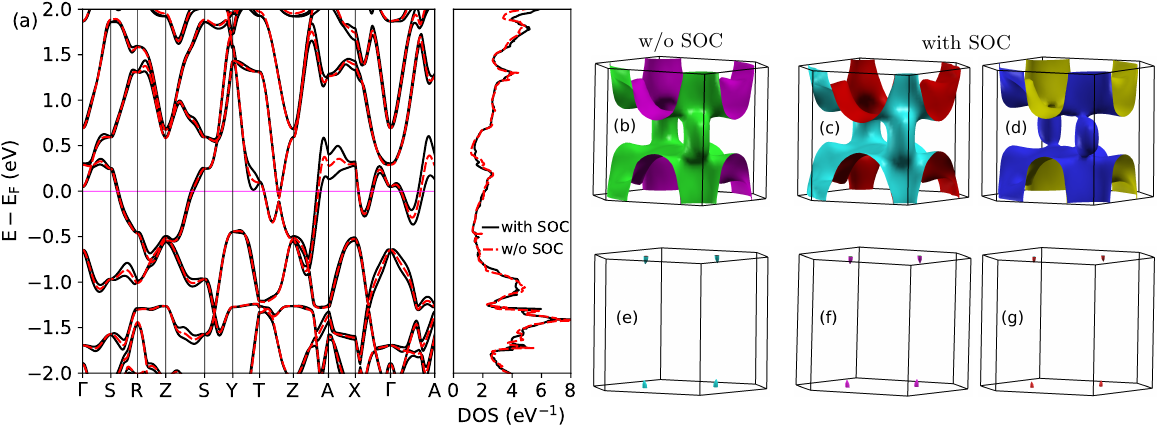}
	\caption{(a) Electronic dispersion relation and density of states of $\mathrm{ThCoC_2}$ calculated with and without SOC. Magenta line marks the Fermi energy. Panels (b) and (e) show Fermi surface sheets without SOC, whereas (c), (d), (f) and (g) present the splitting of the Fermi surface caused by SOC.  
\label{plot_elfs}}
\end{figure*}

{The aim of the present paper is to theoretically investigate the electronic structure, lattice dynamics, electron-phonon interaction and  
superconductivity in 
$\mathrm{ThCoC_2}$, assuming that  superconductivity is  mediated by the electron-phonon interaction. 
In the first step, {\it ab-initio} calculations 
of the electronic band structure and the electron-phonon coupling were performed, and the spin-orbit coupling effects were analyzed.
Next, the thermodynamic parameters of the superconducting state were
determined using the isotropic Eliashberg formalism where all depairing effects were approximated by the Coulomb pseudopotential $\mu^*$. 
We find that already within the Eliashberg framework the superconducting state in $\mathrm{ThCoC_2}$ exhibits non-BCS characteristics.
However, in contrast to $\mathrm{LaNiC_2}$, experimental results strongly deviate both form the BCS predictions and isotropic Eliashberg solutions, showing the importance of the inversion symmetry breaking accompanied by the strong spin-orbit coupling, effects not taken into account in both theories.}

The  paper  is  organized  as  follows.   In  Sec.\ref{sec:comp_details} we provide basic information about $\mathrm{ThCoC_2}$ structure 
and give all the computational details used in the paper. Electronic structure of the considered compound is described in 
Sec.~\ref{sec:electronic_structure} while phonons and electron-phonon coupling are analyzed in Sec.~\ref{sec:eph}. Section \ref{sec:eliashberg} 
contains the analysis of thermodynamic parameters calculated within the Eliashberg formalism, discussing recent experiments in the latter case.
The summary is included in Sec.~\ref{sec:summary}.

\begin{table}[b]
\caption{Experimental and calculated unit cell parameters. Atomic positions (in crystal coordinates) are in a form: Th (0,0,u), Co 
(0.5,0,v), C (0.5,$\pm$y,z).\label{tab_cell}}
\centering
\begin{ruledtabular}
\begin{tabular}{cccccccc}
	 & a (\AA) & b (\AA) & c (\AA) & u & v & y & z\\
	\hline
	expt. & 3.8063 & 4.5329 & 6.1461 & 0.0 & 0.626 & 0.160 & 0.289\\
	calc. & 3.8214 & 4.5376 & 6.0708 & -0.0014 & 0.6041 & 0.1561 & 0.3007\\
	\end{tabular}
\end{ruledtabular}
\end{table}

\section{Computational details\label{sec:comp_details}}

Calculations in this work were done using the Density Functional Theory (DFT) and {\sc Quantum Espresso} package~\cite{Giannozzi2009,QE-2017}.
We used Perdew-Burke-Ernzerhof (PBE) generalized gradient approximation for the exchange-correlation functional~\cite{pbe}
and Rappe-Rabe-Kaxiras-Joannopoulos (RRKJ) ultrasoft pseudopotentials~\cite{ThCoC2pseudos,dalcorso}. 
Calculations were done both in the scalar-relativistic and fully-relativistic approach (that is, with the spin-orbit coupling included). In the latter case, relativistic pseudopotentials for Th and Co were used. 

First, unit cell parameters and atomic positions were optimized 
within Broyden-Fletcher-Goldfarb-Shanno algorithm, starting from the experimental values~\cite{Gerss1986}. Since in the available literature the atomic positions in ThCoC$_2$ are not reported, those of $\mathrm{LaNiC_2}$ were taken as the initial ones. 
Optimized cell parameters are close to the experimental ones and are given in 
Table~\ref{tab_cell}. Atomic positions were additionally relaxed with SOC, but no changes were observed.

For the relaxed unit cell, the electronic structure was calculated using a Monkhorst-Pack grid of $12^3$ {\it k}-points, whereas the Fermi surface was calculated on a denser
$18^3$ mesh. Plane-wave expansion energy and charge density cutoffs were set to $130$~Ry and $1300$~Ry, respectively. Such large values were required due to the presence of the Th atom.
Interatomic force constants (IFC) were computed using the density functional perturbation theory~\cite{dfpt} from the Fourier interpolation of $21$ inequivalent dynamical matrices, which make up a $4^3$ {\it q-}point grid. 
Phonon dispersion relations were calculated from IFC by Fourier interpolation.
Finally, the Eliashberg electron-phonon interaction function $\alpha ^2 F(\omega)$ was determined using the self-consistent 
first order variation of the crystal potential from the preceding phonon calculations.
The spectral function $\alpha ^2 F(\omega)$ was then used to 
determine the electron-phonon coupling (EPC) parameter $\lambda$ and thermodynamic parameters of the  superconducting phase based on the Eliashberg formalism.  

\section{\label{sec:electronic_structure}Electronic structure}

\begin{figure*}[t]
	\includegraphics[width=0.99\textwidth]{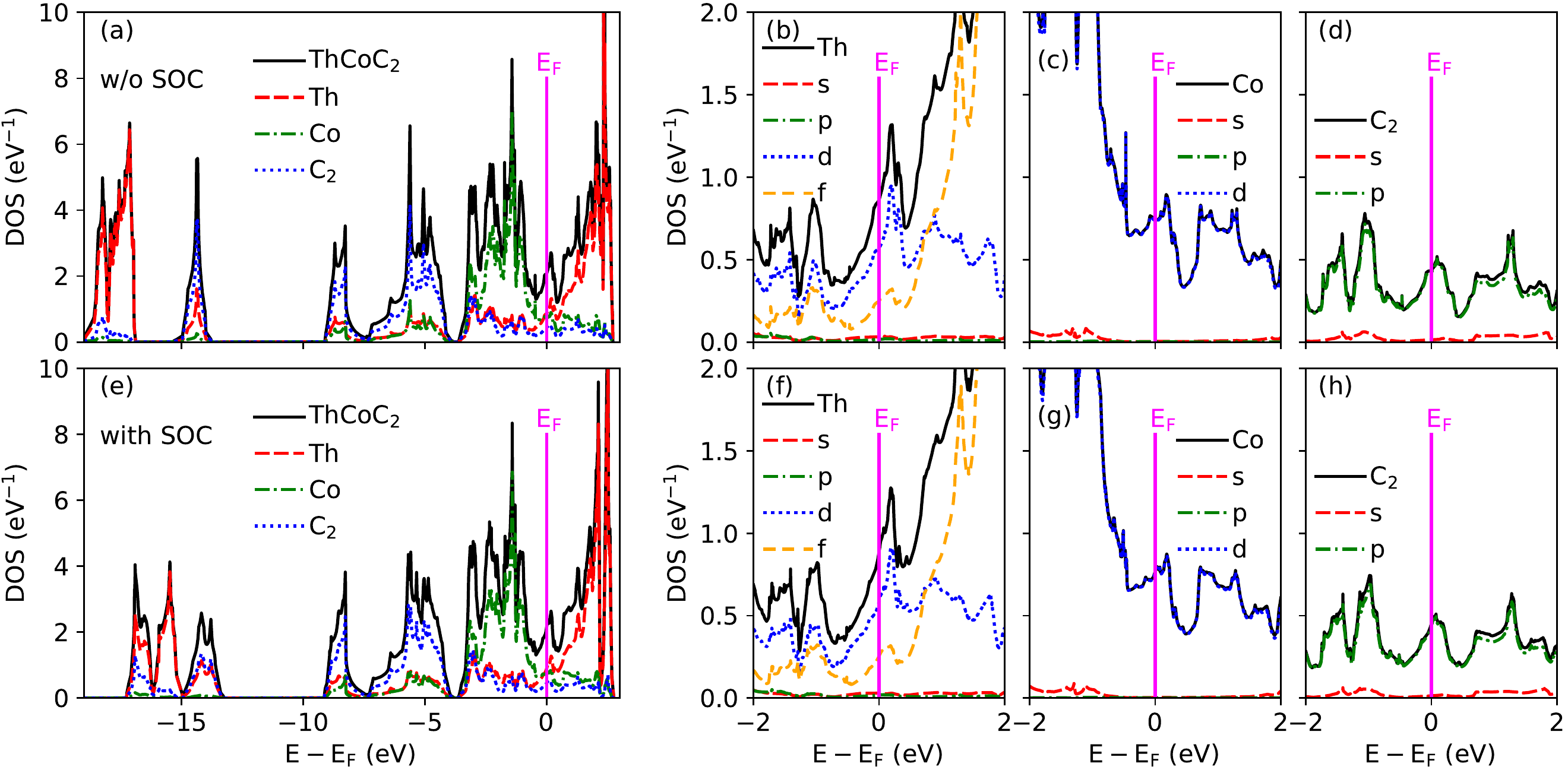}
\caption{Total and projected density of states of $\mathrm{ThCoC_2}$: (a)-(d) without SOC and (e)-(h) with SOC.\label{plot_dos}}
\end{figure*}

Figure \ref{plot_elfs}(a) shows the electronic band structure and density of states (DOS) of $\mathrm{ThCoC_2}$ calculated with and without SOC. Brillouin 
zone of the system, with the location of high-symmetry points, is presented in Fig. \ref{plot_struct}(b). 
As shown, in the scalar-relativistic case, two bands 
cross the Fermi level, forming the Fermi surface (FS) that consists of a complex hole-like sheet [Fig.\ref{plot_elfs}(b)] and two very small electron pockets 
localized on T-Z path in the Brillouin zone [Fig.\ref{plot_elfs}(e)]. 
Interestingly, the hole-like sheet reveals a quasi-two-dimensional topology in the form of two 
parts connected by tubes, all aligned along the $\pmb{k}_z$ axis, what suggests anisotropy in the transport properties. 
Note that our Fermi surface is similar to the one reported by Bhattacharyya et al. \cite{Bhattacharyya2019} even though they employed LDA+U method, which resulted e.g. in removing the small electron pockets from the FS. 

\begin{figure}[b]
	\includegraphics[width=0.50\textwidth]{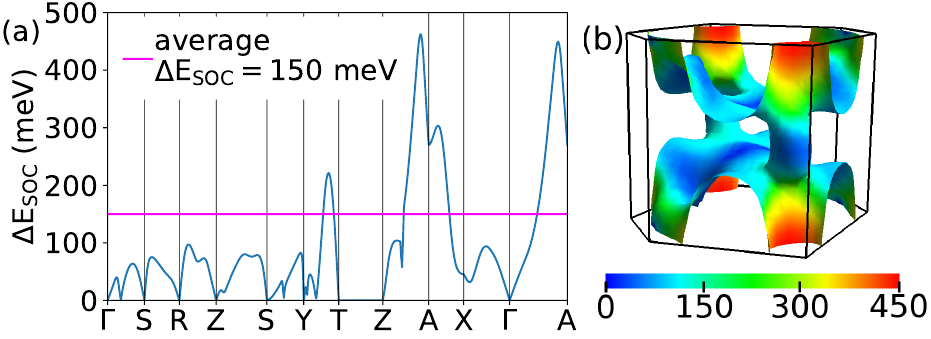}
\caption{$\Delta E_{\rm SOC}$: SOC-induced splitting of the energy band, which gives the dominating contribution to $N(E_F)$: (a) along the high-symmetry directions regardless of the actual energy value; (b) computed under the condition that the two spin-split bands are on the opposite side of the Fermi energy, and marked on the scalar-relativistic Fermi surface, plotted using {\sc FermiSurfer}~\cite{fermisurfer}. The average value of $\overline{\Delta E_{\rm SOC}} = 150$ meV is calculated based on results visualized in panel (b). \label{fig-split}}
\end{figure}

In the relativistic case, the presence of the antisymmetric SOC leads to the splitting of electronic bands.
The two bands crossing the Fermi level split into four 
which results in four spin-split Fermi surface sheets, plotted in Fig.~\ref{plot_elfs}(c,d,f,g). 
Except for the band splitting, SOC does not change the shape of the Fermi surface much,
which looks very similar to that calculated for LaNiC$_2$~\cite{lanic2}.
However, the magnitude of SOC splitting energy $\Delta E_{\rm SOC}$ is considerably larger. $\Delta E_{\rm SOC}$ for the band which forms the dominating FS sheet [Fig.~\ref{plot_elfs}(b-d)], plotted along the same path as used in  
Fig.~\ref{plot_elfs}(a) is shown in Fig.~\ref{fig-split}(a) and reaches its maximal value of $450$~meV near the A point in BZ.

\begin{table}[b]
\caption{Total and projected densities of states at the Fermi energy of $\mathrm{ThCoC_2}$, Sommerfeld coefficient $\gamma$ ("bare" bandstructure values) calculated with and without SOC, and the electron phonon coupling constant $\lambda$, computed as a renormalization factor from the experimental $\gamma_{\rm expt}=\gamma_{\rm band}(1+\lambda)$.
Third row shows the experimental values of $\gamma$ and $\lambda$, the latter is extracted from the experimental $T_c$ and McMillan formula (see text).\label{tab_el_pdos}}
\begin{center}
\begin{ruledtabular}
\begin{tabular}{ccccccc}
         & \multicolumn{4}{c}{$N(E_F)$ (eV$^{-1}$)} & $\gamma$ & $\lambda$\\
		 & Total & Th & Co & $\mathrm{C_2}$ & $\mathrm{\left(\frac{mJ}{mol\, K^2}\right)}$ & \\
		\hline
		w/o SOC & 2.07 & 0.86 & 0.74 & 0.43 & 4.88 & 0.72\\
		with SOC & 2.14 & 0.89 & 0.77 & 0.45 & 5.04 & 0.66\\
		expt. \cite{Grant2017} & \multicolumn{4}{c}{-} & 8.38 & 0.49\\
	\end{tabular}
\end{ruledtabular}
\end{center}
\end{table}

To calculate the average value of SOC splitting, we have computed the energy eigenvalues on a regular grid of about 7000 $k-$points, sampling the primitive cell in the reciprocal space. To get the representative value around the Fermi level only, we have imposed a condition that one of the energy eigenvalues has to be below and the other above $E_F$. About 700 points have met that condition, and the average splitting, computed from the energy differences between the pair of states, {as well as the median,} is about 150 meV.
The distribution of $\Delta E_{\rm SOC}$ in the Brillouin zone is marked on the scalar-relativistic Fermi surface in Fig.~\ref{fig-split}(b).
Computed $\overline{\Delta E_{\rm SOC}}$ is considerably larger than in LaNiC$_2$, where {the average} was about $40$ meV~\cite{lanic2,Smidman_2017}, and closer to that in triplet superconductors CePt$_3$Si and Li$_2$Pt$_3$B (200 meV).  The combination of a large $\Delta E_{\rm SOC}$ and a relatively low $T_c$ results in a large value of the characteristic $E_r$ ratio~\cite{BiPd} $ E_r = \overline{\Delta E_{\rm SOC}}/k_BT_c \simeq 700$. The large value of $E_r$ was observed to be correlated with the presence of a triplet component in several NCSs~\cite{BiPd}. 
$E_r$ is equal to 177 in LaNiC$_2$, $\sim 890$ in Li$_2$Pt$_3$B and $\sim 3000$ in CePt$_3$Si.

Density of states plotted in Fig.~\ref{plot_dos} indicates that the largest contribution to the total DOS at the Fermi level, $N(E_F) = 2.1$ eV$^{-1}$,
comes from $6d$ Th, $3d$ Co and $2p$ C states. 
Due to the charge transfer from the $7s$ shell, $5f$ states of Th are partially filled and contribute to $N(E_F)$ in about 12\%. The Th-$5f$ shell filling is equal about 0.8, similar to that found in elemental Th~\cite{actinides-review} or recently studied ThIr$_3$ superconductor~\cite{thir3}. 
Those $5f$ states have an itinerant character, thus no strong electronic correlations associated with the $f$-shell are expected, 
in agreement with the measured rather low value of the Sommerfeld coefficient (see below).
Along with the small effect on the shape of Fermi surface, SOC changes $N(E_F)$ only slightly (see Table~\ref{tab_el_pdos}). 
The obtained $N(E_F)$ is relatively low and lower than e.g. in LaNiC$_2$ (2.37 eV$^{-1}$~\cite{lanic2}). Resulting band-structure value of the Sommerfeld coefficient $\gamma_{{\rm band}} = \frac{\pi^2}{3}k_B^2N(E_F)$ is
$\gamma_{{\rm band}} = 5.04$ mJ mol$^{-1}$ K$^{-2}$, whereas the
experimental one~\cite{Grant2014} is $\gamma_{\rm expt} = 8.38$ mJ mol$^{-1}$ K$^{-2}$. Under the assumption that the electronic specific heat is renormalized only by the electron-phonon interaction, $\gamma_{\rm expt}=\gamma_{\rm band}(1+\lambda)$, we obtain the estimation of the electron-phonon coupling parameter 
$\lambda = 0.66$, which is in the moderate-coupling regime. 
Table~\ref{tab_el_pdos} also presents the "experimental" value~\cite{Grant2017} of $\lambda = 0.49$, which was recalculated from the McMillan's formula for $T_c$ using the experimental value of $T_c$, Debye temperature of $\theta_D = 449$~K and the Coulomb pseudopotential parameter $\mu^* = 0.13$.

\section{\label{sec:eph}Phonons and electron-phonon coupling}

\begin{figure}[t]
	\centering
	\includegraphics[width=0.99\columnwidth]{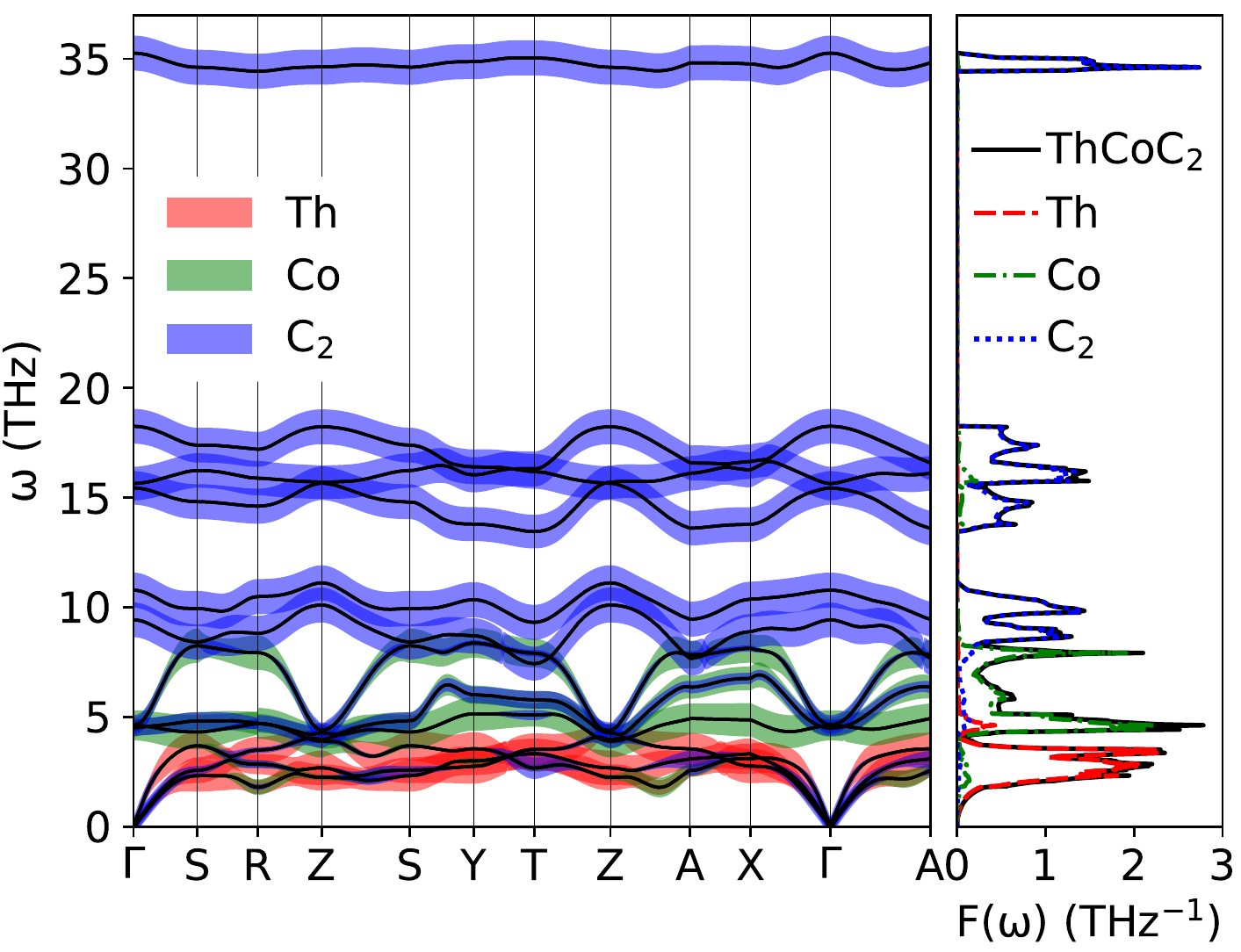}
	\caption{Phonon dispersion relations with colored atomic contributions (left panel) and the corresponding phonon density of states (right 
panel). Results for $\mathrm{ThCoC_2}$ with SOC.}
\label{plot_phbands_char}
\end{figure}

Figure~\ref{plot_phbands_char} presents the phonon dispersion curves 
$\omega(\mathbf{q})$ and the phonon density of states $F(\omega)$ calculated with SOC, although influence of the spin-orbit interaction occurred to be negligible for the phonon structure. 
The obtained phonon spectrum is stable with no imaginary frequencies. 
As there are four atoms in the primitive cell, $12$ phonon branches are displayed in Fig.~\ref{plot_phbands_char}, shaded with respect to the atomic character. The corresponding partial atomic $F(\omega)$ is also presented in the right panel.

Due to the large differences in the atomic masses between Th (232 u), Co (58.9 u), and C (12 u), the phonon spectrum is separated into three parts. 
The six highest frequency phonon branches are dominated by carbon atoms' vibrations, with the single mode near $35$~THz attributed to the bond-stretching C-C oscillations, just like observed in LaNiC$_2$~\cite{lanic2}. 
Cobalt dominates among the modes located between 5 and 10 THz, whereas the lowest part of the spectrum is mostly contributed by thorium. 
The average phonon frequencies are shown in Table~\ref{tab:avph} and one can see that SOC only slightly lowers the average values.

\begin{table}[b]
\caption{Average phonon frequencies. \label{tab:avph}}
\begin{center}
\begin{ruledtabular}
\begin{tabular}{ccccc}
		 & Total (THz) & Th (THz) & Co (THz) & C (THz) \\
		\hline
		with SOC & 10.59 & 3.00 & 6.41 & 16.55 \\
		w/o SOC & 10.62 & 3.00 & 6.42 & 16.60 \\
	\end{tabular}
\end{ruledtabular}
\end{center}
\end{table}

Electron-phonon interaction is described within the isotropic Eliashberg function, defined as
\begin{equation}
	\alpha^2F(\omega) = \frac{1}{2\pi N(E_F)} \sum_{\pmb{q}\nu} \delta(\omega - \omega_{\pmb{q}\nu}) \frac{\gamma_{\pmb{q}\nu}}{\hbar 
\omega_{\pmb{q}\nu}}, \label{eq_a2F}
\end{equation}
where $\omega_{\pmb{q}\nu}$ is the phonon frequency for the mode $\nu$ at the wave vector $\pmb{q}$ while $\gamma_{\pmb{q}\nu}$ is the 
phonon linewidth 
which describes the strength of the interaction between the electrons from the Fermi surface and a phonon 
mode $\nu$ at $\pmb{q}$:
\begin{eqnarray}
	\gamma_{\pmb{q}\nu} = 2\pi \omega_{\pmb{q}\nu} \sum_{ij} &\int& \frac{d^3k}{\Omega_{BZ}}  |g_{\pmb{q}\nu}(\pmb{k},i,j)|^2 \nonumber \\
&\times& \delta(E_{\pmb{q},i} - E_F) \delta(E_{\pmb{k}+\pmb{q},j} - E_F). \label{eq_gammaq}
\end{eqnarray}
In the above expression, $g_{\pmb{q}\nu}(\pmb{k},i,j)$ are the interaction matrix elements given by \cite{Heid2010} 
\begin{equation}
 g_{\pmb{q}\nu}(\pmb{k},i,j)=\sum_s \left ( \frac{\hbar}{2M_s\omega_{\mathbf{q}\nu}} \right ) ^{1/2} \langle \psi _{i,\mathbf{k}}| \frac{dV_{scf}}{d\hat{u}_{\nu,s}} \cdot \hat{\epsilon}_\nu | \psi _{j,\mathbf{k+q}} \rangle,
\end{equation}
where $M_s$ is a mass of the atom $s$, $\psi_{i,\mathbf{k}}$ is an electron wave function at $\mathbf{k}$-point, $\hat{\epsilon}_\nu$
is a polarization vector of a phonon mode and $\frac{dV_{scf}}{d\hat{u}_{\nu,s}}$ is a change of the electronic potential, calculated in the 
self-consistent manner, due to the displacement of the atom $s$ in the $\hat{u}_{\nu,s}$ direction.
While the Eliashberg function expresses the electron-phonon coupling summed over Brillouin zone and phonon modes $\nu$, the electron-phonon coupling 
constant (EPC) $\lambda$ is a single parameter characterizing the overall coupling strength, and is given by	
\begin{equation}
	\lambda = 2 \int_0^{\omega_{max}} \frac{\alpha^2F(\omega)}{\omega} d\omega. \label{eq_lambda_a2F}
\end{equation}

\begin{figure}[t]
	\centering
	\includegraphics[width=0.99\columnwidth]{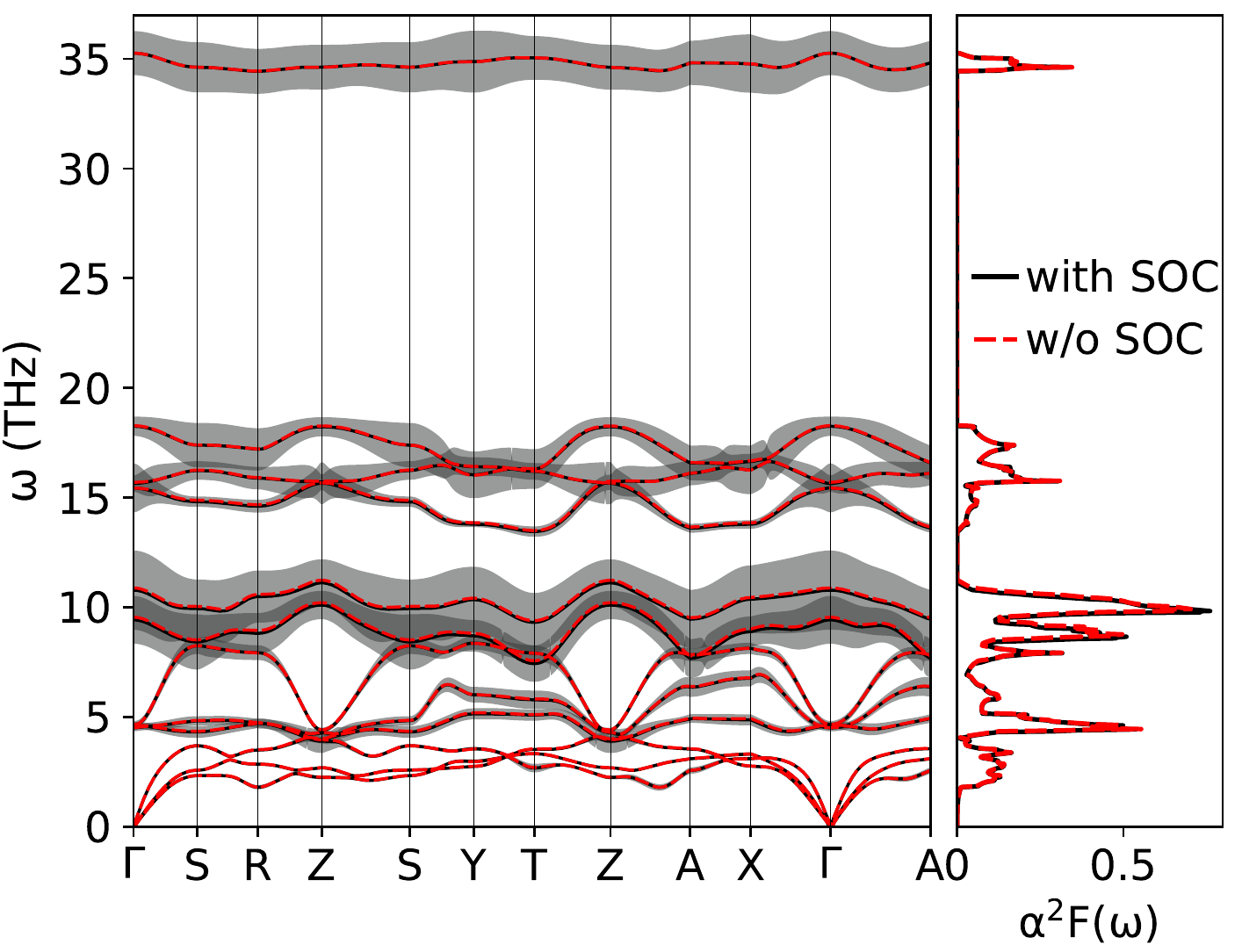}
	\caption{Electron-phonon coupling in $\mathrm{ThCoC_2}$. Left panel: phonon dispersion relations with the phonon linewidths $\gamma_{\pmb{q}\nu}$ 
marked by shading. Phonon linewidths are multiplied by 20. Right panel: Eliashberg function computed with and without SOC.\label{plot_elph}}
\end{figure}

\begin{figure}[t]
	\centering
	\includegraphics[width=0.99\columnwidth]{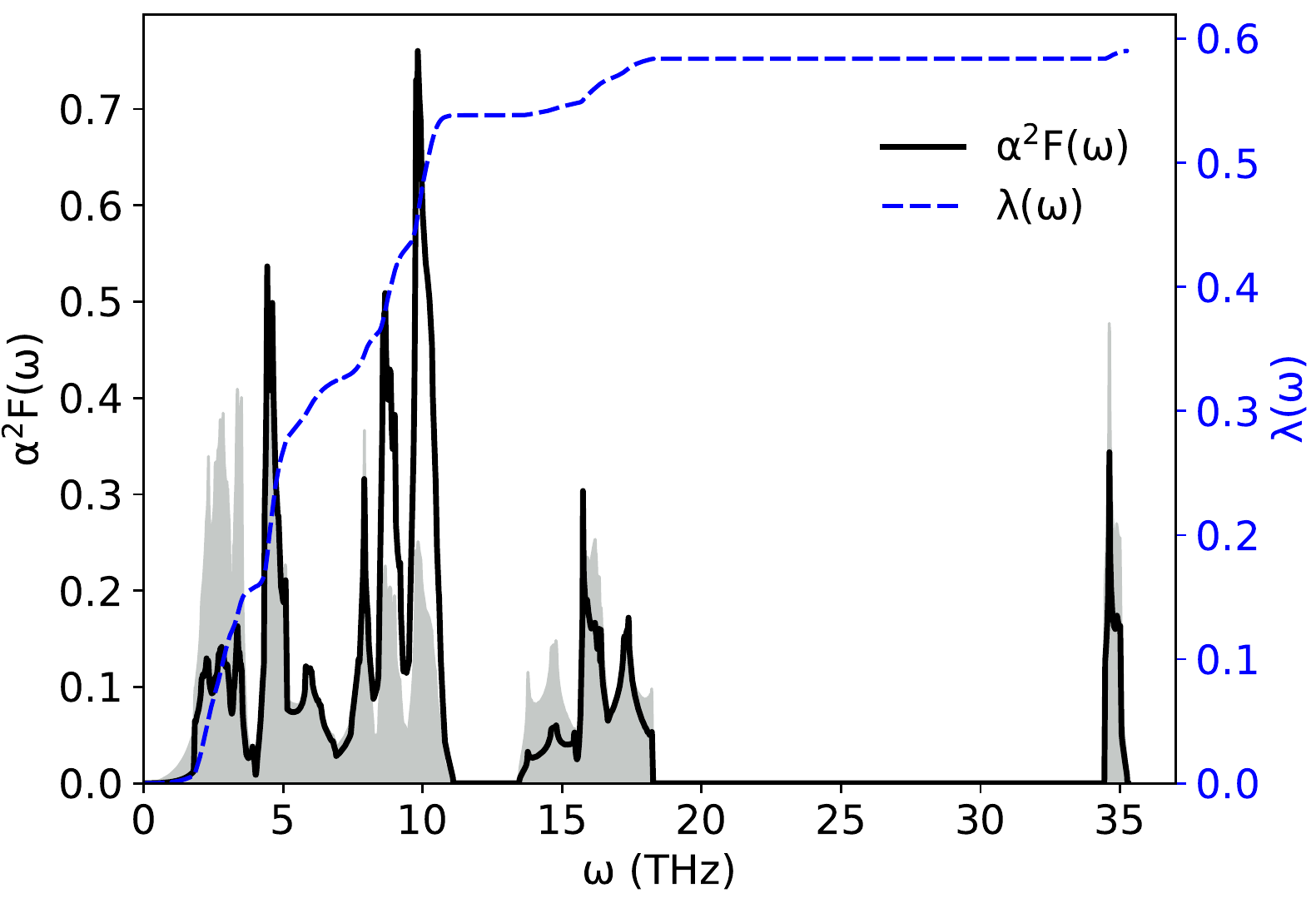}
	\caption{Eliashberg function of $\mathrm{ThCoC_2}$ and cumulative electron-phonon coupling constant. Shading displays phonon density of states normalized to the integral of the Eliashberg function. Results for the case with SOC. \label{plot_lambc}}
\end{figure}

Figure~\ref{plot_elph} displays the phonon dispersion curves with shading corresponding to the phonon linewidths, $\gamma_{\pmb{q}\nu}$, together with 
the Eliashberg spectral function presented in the right panel. 
As we mentioned above, SOC had almost no effect on the phonon dispersion curves, the same holds for the magnitude of the electron-phonon coupling, with no visible differences for the scalar-relativistic and relativistic $\alpha^2F(\omega)$. 
All these are a consequence of the fact that lighter atoms'  (Co and C) electronic states  give the largest contribution to the Fermi surface, and the contribution from Th is not a dominant one. Much different situation was found in e.g. CaBi$_2$~\cite{cabi2} or LaBi$_3$~\cite{labi3}, 
where due to domination of Bi-$6p$ orbitals the electron-phonon coupling constant was modified by SOC in almost 50\%.
The computed linewidths $\gamma_{\pmb{q}\nu}$ show that the electron-phonon coupling is relatively the strongest for the carbon atoms' modes,
located around $10$~THz. The highest $C$ mode, located near $35$~Hz, despite having large $\gamma_{\pmb{q}\nu}$, does 
not give a significant contribution to the electron-phonon coupling as $\alpha^2F(\omega)$ is inversely proportional to $\omega$, and $\lambda \propto \frac{\gamma_{\pmb{q}\nu}}{\omega_{\pmb{q}\nu}^2}$, see Eq.(\ref{eq_a2F}-\ref{eq_lambda_a2F}).
Additionally, in Fig.~\ref{plot_lambc} Eliashberg function is plotted with the phonon DOS in the background and the cumulative electron-phonon coupling parameter $\lambda(\omega)$.
Here one can clearly see the relative enhancement of $\alpha^2F(\omega)$ above the bare $F(\omega)$ near 10 THz due to the large $\gamma_{\pmb{q}\nu}$ of carbon.
As presented in Fig.~\ref{plot_elph} and Fig.~\ref{plot_lambc}, the meaningful contribution to the electron-phonon coupling comes also from the Co vibrations. Although Co modes 
couple to electrons to a smaller extent, having a significantly narrower linewidth, 
the lower frequencies in its phonon branches partly compensate smaller $\gamma_{\pmb{q}\nu}$  giving raise to a large peak in $\alpha^2F(\omega)$ near $5$~THz. 
Relatively, the weakest coupling is seen for Th modes, for which Eliashberg function goes visibly below $F(\omega)$ in Fig.~\ref{plot_lambc}, nevertheless due to their lowest phonon frequencies thorium modes do contribute to the total $\lambda$. 
As one can see from the cumulative $\lambda(\omega)$ plot, the electron phonon coupling constant is determined mostly by the phonon modes from 0 to 10~THz with important contributions from Th, Co and the two lowest C modes.  
The calculated total EPC parameter is $\lambda = 0.59$ and is only slightly enhanced by SOC, see Table \ref{tab_lambda_Tc}.
This value is close to $\lambda = 0.49$, determined from $T_c$ using the inverted McMillan formula in Ref.~\cite{Grant2017}, and is slightly lower than extracted above as a renormalization factor of the Sommerfeld coefficient, $\lambda = 0.66$. All these values suggest that ThCoC$_2$ is in the intermediate electron-phonon coupling regime. 
In Table \ref{tab_lambda_Tc} we also include the computed values of the renormalized Sommerfeld coefficient 
$\gamma = \gamma_{\rm band}(1+\lambda)$, where $\gamma_{\rm band}$ were determined from the computed DOS in Table~\ref{tab_el_pdos}, and $\lambda$ from the Eliashberg function.
The theoretical value of 8.0~$\mathrm{(mJ \ mol^{-1} \ K^{-2})}$ is slightly lower than the experimental one, which is 8.4~$\mathrm{(mJ \ mol^{-1} \ K^{-2})}$.

\begin{figure}[t]
	\centering
	\includegraphics[width=0.48\textwidth]{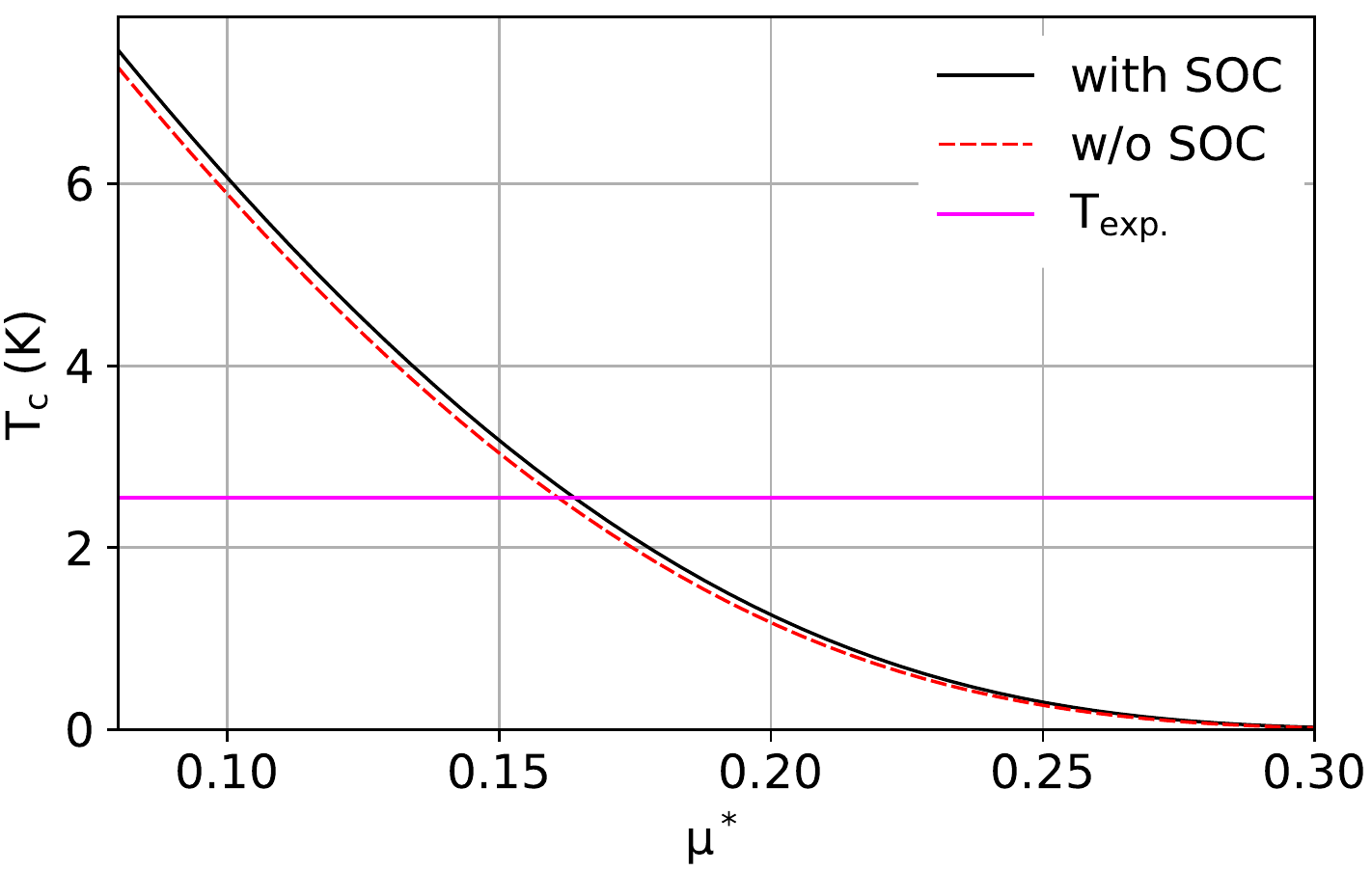}
	\caption{Critical temperature of $\mathrm{ThCoC_2}$ calculated as a function of the $\mu^*$ parameter from the Allen-Dynes formula. Magenta line marks the experimental value $T_c=2.55$~K. \label{plot_tcmu}}
\end{figure}

Next, we may calculate the superconducting critical temperature  which would correspond to the calculated $\lambda$ under the assumption of the conventional $s$-wave superconducting gap symmetry. {Although for the noncentrosymmetric compounds the Allen-Dynes formula is used beyond its applicability, it is commonly applied to estimate $T_c$ while the value of $\mu^*$ parameter provides an additional information about the degree of deviation of superconductivity from the conventional one as well as about the strength of electronic interactions.}
The Allen-Dynes formula~\cite{allen} has the following form 
\begin{equation}\label{eq_Allen_Dynes}
k_{B}T_{c}=\frac{\hbar\langle\omega_{\rm log}^{\alpha^2F}\rangle}{1.20}\,
\exp\left\{-\frac{1.04(1+\lambda)}{\lambda-\mu^{*}(1+0.62\lambda)}\right\},
\end{equation}
where $\mu^*$ is the Coulomb pseudopotential parameter and 
$\langle\omega_{\rm log}^{\alpha^2F}\rangle$ is the logarithmic average
\begin{equation}
	\langle\omega_{\rm log}^{\alpha^2F}\rangle = \exp \left( \frac{2}{\lambda}\int_0^{\omega_{max}} \alpha^2F(\omega) \ln \omega \frac{d\omega}{\omega} %\middle/ 
%\int_0^{\omega_{max}}\alpha^2F(\omega) \frac{d\omega}{\omega}  
\right).
\label{eq_omloga2F}
\end{equation}

$T_c$ as a function of the Coulomb pseudopotential $\mu^*$ is plotted in Fig.~\ref{plot_tcmu}. 
For the typically used values of $\mu^*$ ($0.10$ recommended for use with the Allen-Dynes formula~\cite{allen} or 0.13 commonly-used with the McMillan equation) $T_c$ is overestimated roughly two times.  There are several possible explanations for this overestimating tendency.  The first is that the electron-phonon coupling is weaker in the 
real material than the calculated $\lambda$. However, as the computed renormalized Sommerfeld parameter is lower than the experimental one (8.0 {\it versus} 8.4 mJ mol$^{-1}$ K$^{-2}$), this is not expected. The second is that the Coulomb pseudopotential $\mu^*$ is  enhanced. 
The experimental value of $T_c = 2.5$~K is reproduced for $\mu^* = 0.165$, a value larger than typically observed in intermetallic superconductors.
As frequently considered in such cases, this may originate from the presence of other depairing mechanisms which for the considered compound can be the electron-paramagnon interactions (spin fluctuations)~\cite{berk,winter79,paramagnons,mgcni3-paramagnons,mosb-cc,mosb,thir3}, recently suggested as the pairing mechanism for ThCoC$_2$~\cite{Bhattacharyya2019}.  Assuming spin fluctuations as an additional renormalization factor for the electronic specific heat, $\gamma_{\rm expt} = \gamma_{\rm band}(1+ \lambda + \lambda_{\rm sf})$, a small value of $\lambda_{\rm sf} = 0.07$ is obtained.
This hypothesis, however, requires further experimental analysis by measurements of the electrical resistivity and magnetic susceptibility.
It is worth recalling that for Nb and its alloys values of $\mu^*$ as large as $\sim 0.20$ has to be used to reproduce the experimental $T_c$~\cite{savrasov_1996,hea-pressure}. 
Finally, the last and the most probable reason for the need of applying enhanced $\mu^*$ is the 
inapplicability of the Allen-Dynes formula for the ThCoC$_2$ case. 
The strong spin-orbit interaction, leading to the average band splitting of 150 meV, is comparable to the maximal phonon energy (35 THz $\simeq 145$ meV) and larger than the average phonon energy of 44 meV, with $\overline{\Delta E_{\rm SOC}}/k_BT_c \simeq 700$.
Thus, it will effectively compete with the formation of the conventional superconducting phase, as in the limit of the strong spin-orbit interaction~\cite{mixedstate2} the pairing inside the spin-split bands requires the odd parity  with respect to the $\mathbf{k}\rightarrow -\mathbf{k}$.
As a consequence, the Allen-Dynes formula, derived for a conventional case, may overestimate $T_c$,
which was not the case for LaNiC$_2$, where experimental $T_c$ was obtained for $\mu^*=0.13$~\cite{lanic2,lanic2-singh}.
However, the electron-phonon interaction seems to be strong enough to be responsible for superconductivity in $\mathrm{ThCoC_2}$ with the observed $T_c$, supporting the idea of the conventional pairing mechanism, but the analysis of the thermodynamic properties of the superconducting state, presented in the next Section, reveals 
{further deviations from the simplest isotropic $s$-wave picture.}

\begin{table}[t]
	\caption{Logarithmic average $\langle\omega_{\rm log}^{\alpha^2F}\rangle$, electron-phonon coupling constant $\lambda$, superconducting critical 
temperature $T_c$ from eq. (\ref{eq_Allen_Dynes}) with $\mu^*=0.13$, and Sommerfeld coefficient $\gamma$ renormalized with 
$\lambda$ from the electron-phonon 
calculations, $\gamma = \gamma_{\rm band}(1+\lambda)$. Experimental values repeated for convenience.\label{tab_lambda_Tc}}
\begin{center}
\begin{ruledtabular}
\begin{tabular}{ccccc}
		 & $\langle\omega_{\rm log}^{\alpha^2F}\rangle$ & $\lambda$ & $T_c$ & $\gamma$\\
		 & (THz) & & (K)  & $\mathrm{\left(\frac{mJ}{mol\, K^2}\right)}$\\ %$\mathrm{(mJ \ mol^{-1} \ K^{-2})}$\\
		\hline
		w/o SOC & 5.849 & 0.583 &  4.05& 7.73\\
		with SOC & 5.816 & 0.590 & 4.23& 8.01\\
		expt. \cite{Grant2017} & - & 0.493 & {2.55} & 8.38\\
	\end{tabular}
\end{ruledtabular}
\end{center}
\end{table}

\section{ELIASHBERG FORMALISM}
\label{sec:eliashberg}
{
In this section, we determine thermodynamic parameters of $\mathrm{ThCoC_2}$ in the superconducting state within the isotropic  Eliashberg formalism~\cite{Eliashberg1960}, and compare 
our results with the latest experiments. 
The usage of the Eliashberg equations is considered as a first approximation towards the understanding of superconductivity in ThCoC$_2$, since similarly to the Allen-Dynes formula, they are not fully applicable for the case of a noncentrosymmetric compound with strong SOC. 
However, comparing the results obtained using this approach with the experiment allows to directly show how strong are the deviations of
superconducting properties of ThCoC$_2$ from the conventional isotropic ones.
}

\subsection{Eliashberg equations}
Our calculations are based on the isotropic Eliashberg equations which, on the imaginary axis ($i=\sqrt{-1}$), are given by
\begin{eqnarray}
Z(i\omega _n)&=&1+\frac{\pi k_B T}{\omega _n} \sum _{n'} \frac{\omega _{n'}}{R(i\omega _{n'})} \lambda(n-n')
 \label{ee1} \\
 Z(i\omega _n) \Delta(i\omega _n) &=& \pi k_B T \sum _{n'} \frac{\Delta(i\omega _{n'})}{R(i\omega _{n'})} \times \nonumber \\
 &&[ \lambda(n-n')-\mu^* \theta ( \omega _c - \omega _{n'} )],
 \label{ee2}
\end{eqnarray}
where $Z(i\omega_n)$ is the mass renormalization function, $\Delta(i\omega _n)$ is the superconducting order parameter, $i\omega_n=i(2n+1)\pi k_B T$ 
are fermionic Matsubara frequencies where $n\in \mathbb{Z}$, $\theta(\omega)$ is the Heviside function, $k_B$ is the Boltzmann constant, $T$ is  
temperature and $R(i\omega _n) = \sqrt{\omega _n ^2 + \Delta^2(i\omega _n)}$. 

The kernel of the electron-phonon interaction is assumed in the common form 
\begin{equation}
 \lambda(n-n')=\int _0^{\infty} d \omega \frac{2 \omega \alpha ^2 F(\omega)}{(\omega _n - \omega _{n'})^2+\omega ^2},
\end{equation}
where $\alpha ^2 F(\omega)$ is the isotropic Eliashberg spectral function discussed in Sec.~\ref{sec:eph}.

In Eq.~(\ref{ee2}), $\mu ^*$ is the Coulomb pseudopotential which is originally defined as a double Fermi surface average of the matrix elements of the screened Coulomb interactions between electrons taking part in the scattering events $\mathbf{k} \rightarrow \mathbf{k}'$ induced by the electron-electron interaction. $\mu^*$ usually yields the value in the  range $[0.1,0.2]$~\cite{Morel1962}. Note however that in some electron-phonon mediated superconductors, $\mu^*>0.2$, which cannot result solely from the Coulomb interaction. Then, as commonly assumed, $\mu ^*$ contains all physical effects competitive to superconductivity and not included in the Eliashberg equations.
Although calculations of $\mu^*$ from the {\it ab-initio} methods are possible, it requires more sophisticated numerical methods~\cite{pickett,mgb2-mu,gross} which are beyond the scope of the present paper. Instead of that, we use the common practice in which $\mu^*$ is determined based on the experimental value of $T_c$ to be able to discuss the thermodynamic properties of the material as a function of $T/T_c$.
Since it corresponds to the Coulomb pseudopotential in the Allen-Dynes formula, they are denoted by the same symbol, although usually different values are required to get the same $T_c$ in both approaches~\cite{lanic2,szczesniak_2015,srm2}, {and a scaling procedure should be applied~\cite{allen} to compare both values.}

The isotropic Eliashberg equations (\ref{ee1})-(\ref{ee2}) are solved iteratively until the convergence is reached, which we consider to occur when 
the relative variation of $\Delta(i\omega_n)$ between two consecutive iterations is lower than $10^{-9}$. The number of iterations is reduced by 
the use of the Broyden method to predict subsequent solutions~\cite{Broyden1984}. The calculations are performed for the cut-off frequency 
$\omega _c = 4 \omega _{max}$~\cite{Carbotte1990} and the number of Matsubara frequencies $M=6500$. The self-consistent solution of 
Eqs.~(\ref{ee1})-(\ref{ee2}) for a given Eliashberg spectral function $\alpha^2F(\omega)$ is then used to calculate thermodynamic parameters and  compare them with experiments.

\subsection{Results}
\begin{figure}[t]
\includegraphics[scale=.6]{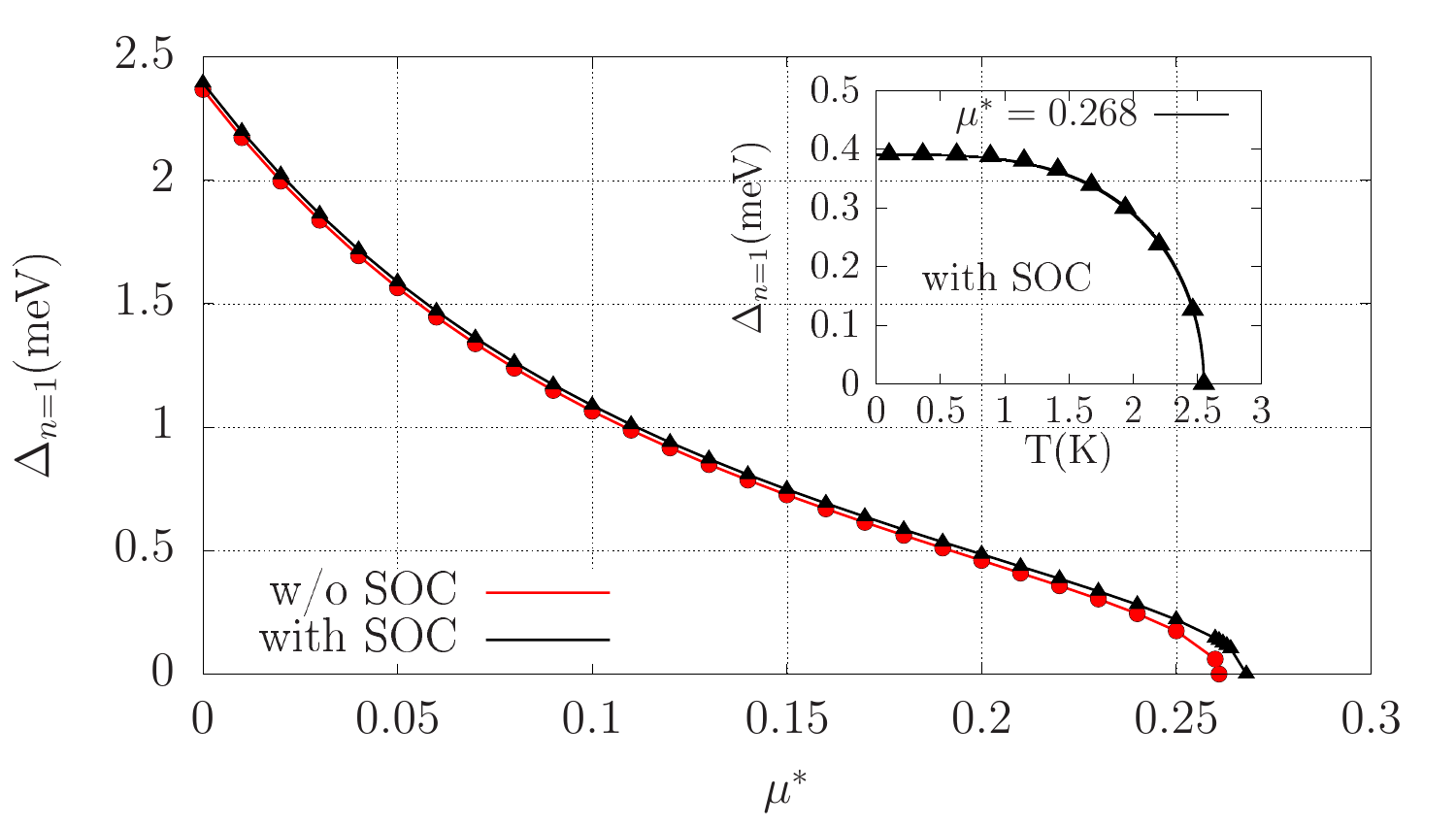}
\caption{ The superconducting energy gap $\Delta_{n=1}$ evaluated at $T=2.55$~K, as a function of $\mu^*$. Results with and without SOC. Inset 
presents $\Delta_{n=1}(T)$ for $\mu^*=0.268$ for which $T_c=2.55$~K. }
\label{fig:muTc}
\end{figure}
To determine $\mu^*$ corresponding to the experimental $T_c=2.55$~K,  we solved the equation $\Delta_{n=1}(T=2.55$ K$)=0$ for different $\mu^*$, see  Fig.~\ref{fig:muTc}. The value of $\mu^*$ for which $\Delta_{n=1}(T=2.55$ K$)=0$ defines the effective Coulomb parameter which should be taken  into account to correctly evaluate thermodynamic parameters of ThCoC$_2$. As shown in Fig.~\ref{fig:muTc}, in the case with SOC, this 
procedure gives a relatively large value of $\mu ^*=0.268$, larger than 0.22 required to get the experimental value of $T_c$ from Eliashberg formalism in LaNiC$_2$~\cite{lanic2}.
{To be compared with $\mu^*$ used with the $T_c$ equation (\ref{eq_Allen_Dynes}) it should be scaled to $\widetilde{\mu^*}$ according to the equation~\cite{allen}
\begin{equation}
\frac{1}{\widetilde{\mu^*}} = \frac{1}{\mu^*} + \ln\left(\frac{\omega_c}{\omega_{\rm max}}\right),
\end{equation}
which with the cut-off frequency of $\omega _c = 4 \omega _{max}$ gives $\widetilde{\mu^*} = 0.195$, again larger than typically used $\mu^*$ values, similarly when the Allen-Dynes formula was applied.}

The temperature dependence of $\Delta_{n=1}$ for $\mu^{*}=0.268$ is presented in the 
inset of Fig.~\ref{fig:muTc} and undergoes the following formula
\begin{equation}
 \Delta(T)=\Delta(0)\sqrt{1-\left ( \frac{T}{T_c} \right )^\Gamma},
 \label{eq:delT}
\end{equation}
with $\Gamma=3.31\pm 0.01$ slightly larger than predicted from the BCS theory, $\Gamma_{BCS} \approx 3.0$. The extrapolated 
$\Delta(0)=0.39$~meV gives the dimensionless ratio $R_{\Delta} = 2 \Delta (0) / k_B T_c=3.55$ close to the BCS value $3.53$. If not stated otherwise, 
in the further part of this work we present results with SOC.

\begin{figure}[t]
\includegraphics[scale=.6]{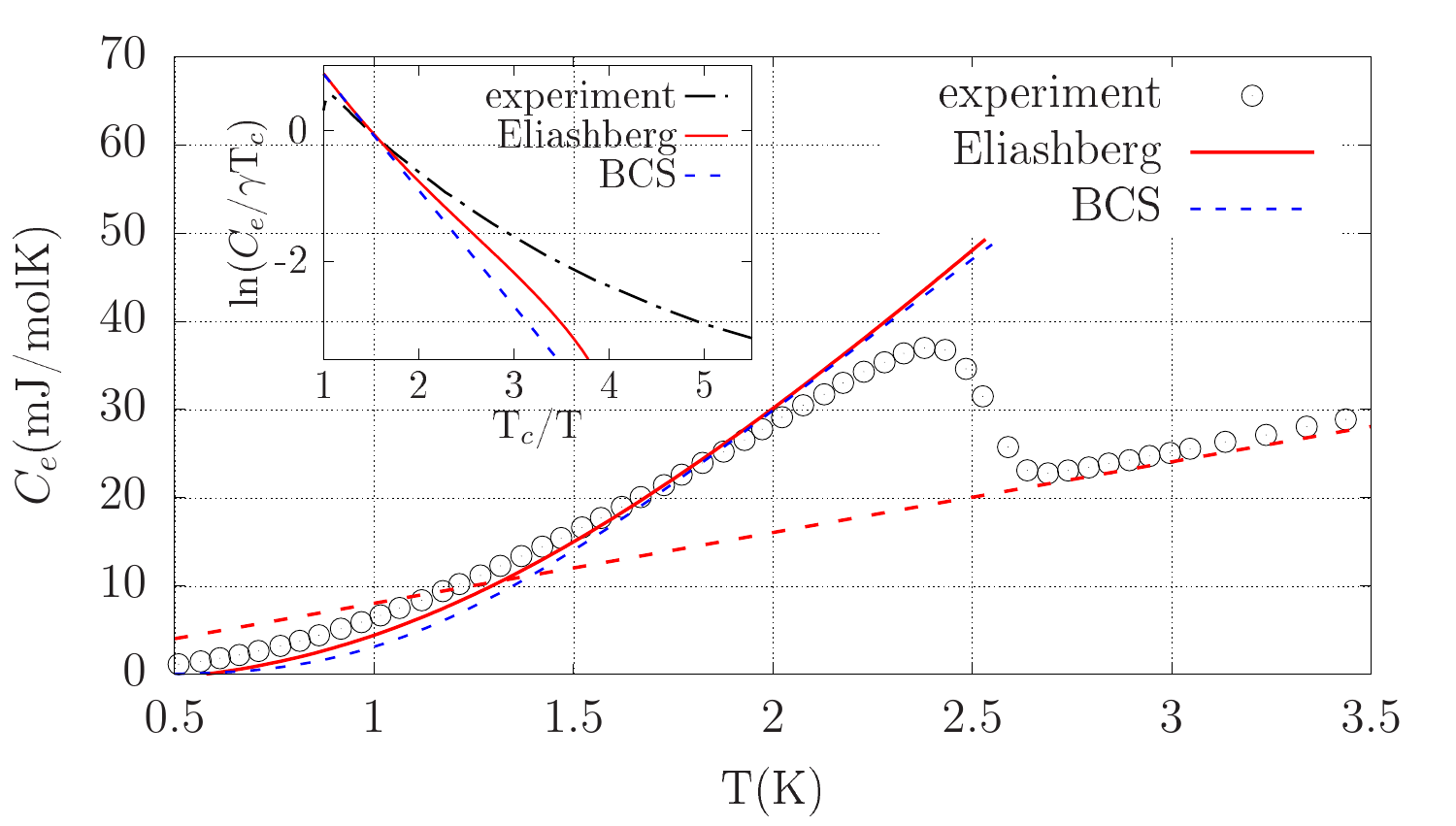}
\caption{Electronic specific heat $C_e$ as a function of temperature $T$.  The normal state is marked by dashed red line while the experimental data~\cite{Grant2014} are marked by dots. Inset  presents the normalized specific heat as a function of the inverse reduced temperature $T/T_c$. }
\label{fig:C}
\end{figure}

For an interacting electron-phonon system, the energy difference between the superconducting and normal state $\Delta F$ is given by
\begin{eqnarray}
 \frac{\Delta F}{N(E_F)}&=&-\pi k_BT \sum _{n} \left ( \sqrt{\omega_n^2+\Delta _n^2} - |\omega _n|\right ) \times \nonumber \\
 && \left ( Z^S(i\omega _n) - Z^N(i\omega _n) \frac{|\omega _n|}{\sqrt{\omega _n^2+\Delta_n^2}} \right ),
\end{eqnarray}
where $N(E_F)$ corresponds to the density of states at the Fermi level while $Z^S$ and $Z^N$ denote the mass renormalization factors 
for the superconducting (S) and normal (N) states, respectively. Then, the difference in the electronic specific heat $\Delta C_e=C_e^S-C_e^N$ can be expressed as
\begin{equation}
 \frac{\Delta C_e(T)}{k_B N(E_F)} = -\frac{1}{\beta}\frac{d^2 \Delta F /N(E_F)}{d(k_BT)^2},
\end{equation}
with the specific heat in the normal state given by
\begin{equation}
\frac{C_e^N(T)}{k_B N(E_F)}=\frac{\pi^2}{3} k_B T (1+\lambda),
\end{equation}
where $\lambda$ is the electron-phonon coupling constant.

The temperature dependence of the specific heat is presented in Fig.~\ref{fig:C} with the experimental data from Ref.~\cite{Grant2014} plotted by dots. For comparison, BCS results are also displayed.  The BCS theory of superconductivity predicts the exponential behavior of the electronic specific heat at low temperatures, in the form 
$C_e \propto \exp[-\Delta(0)/k_BT]$. As shown in the inset of Fig.~\ref{fig:C}, which displays the logarithmic graph of the electronic specific heat versus 
the inverse reduced temperature ($T_c/T$), the experimental curve deviates from the linear plot, expected for weakly coupled superconductors.  Interestingly, at low temperatures, also the Eliashberg solution deviates from the 
linear BCS behavior.
This demonstrates that the non-BCS behavior of $C_e$ may occur even for the isotropic $s$-wave gap symmetry,  if  the retardation effects in the electron-phonon interactions are included as done in the  Eliashberg formalism~\cite{Durajski-h2s}. 
The deviation, however, is not as strong as observed experimentally~\cite{Grant2014}.
At higher temperatures the Eliashberg solution approaches the BCS behavior and again deviates from the experimental data  reaching  the jump of reduced specific heat at $T_c$, $\Delta C_e/\gamma T_c=1.425$, higher than the experimental $0.86$, and almost equal to the weak-coupling BCS limit of $1.43$. 

\begin{figure}[t]
\includegraphics[scale=.6]{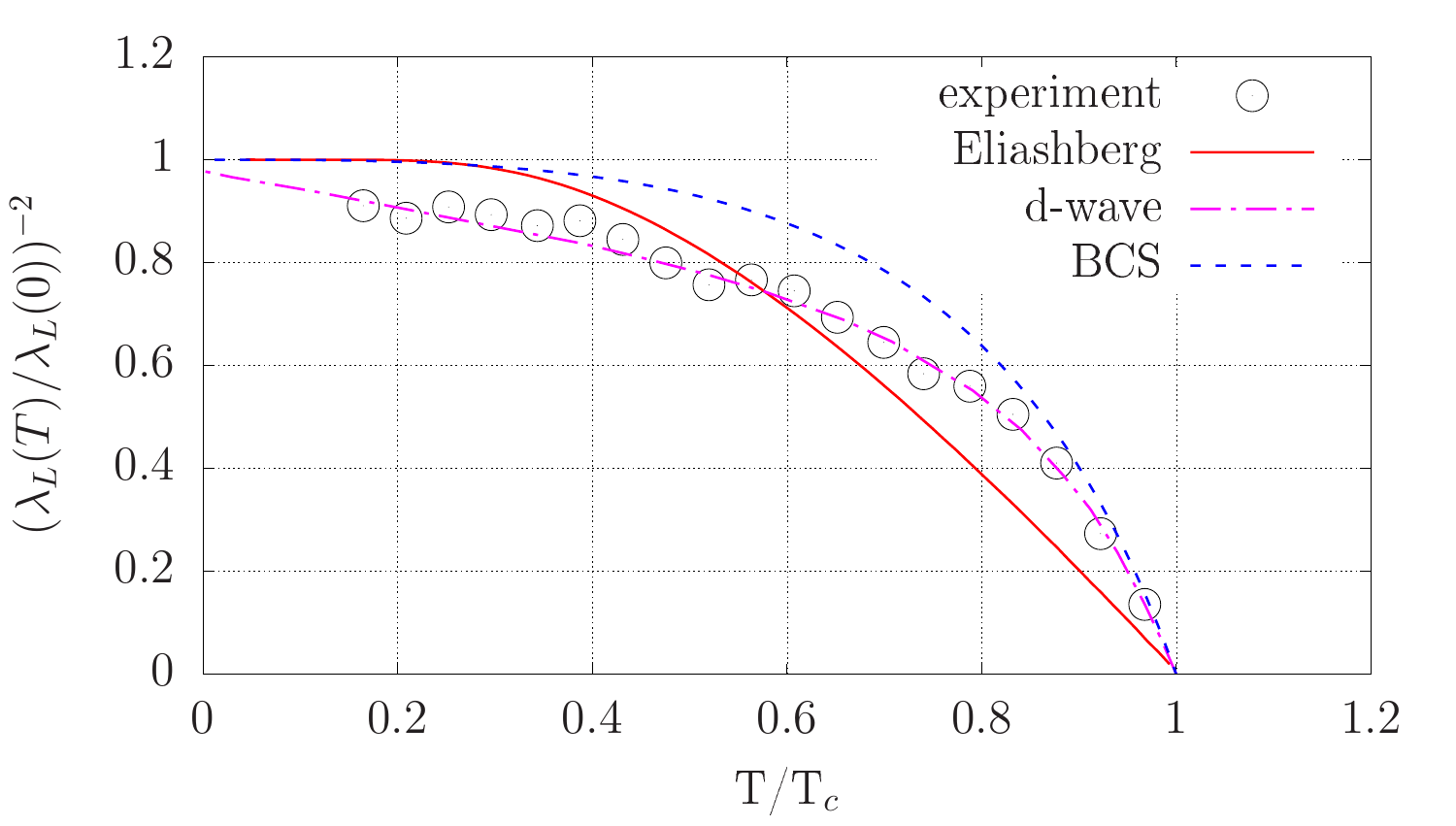}
\caption{Temperature dependence of the inverse normalized magnetic penetration depth $\lambda_L^{-2}$. Experimental data from the $\mu$SR measurements, marked with dots, as well as the $d$-wave fit are taken from Ref.~\onlinecite{Bhattacharyya2019}.}
\label{fig:L}
\end{figure}

Although as presented above, the deviation from a purely exponential behavior in the electronic specific heat measurement does not prejudge the nodal line superconductivity in $\mathrm{ThCoC_2}$, recent $\mu$SR measurements~\cite{Bhattacharyya2019} of the temperature dependence of the magnetic field penetration depth gave a strong support to the nodal superconductivity hypothesis, where the $d$-wave gap symmetry was proposed. 
Within the Eliashberg model, the London penetration depth $\lambda _L$  can be calculated from the expression
\begin{equation}
 \frac{1}{e^2 v_F^2N(E_F)\lambda _L^2(T)}= \frac{2}{3} \pi k_B T \sum _n \frac{\Delta _n^2}{Z^S(i\omega _n)[\omega _n^2+\Delta _n^2]^{3/2}},
\end{equation}
where $e$ is the electron charge and $v_F$ is the Fermi velocity~\cite{Carbotte1990}. 

Figure~\ref{fig:L} displays the comparison of the temperature dependence of the normalized penetration depth $\lambda_L^{-2}$ from the $\mu$SR measurements, $d$-wave fit (both from~\cite{Bhattacharyya2019}), our calculations~\footnote{For calculations presented in Fig.~\ref{fig:L} to keep the consistency with the experimental results, we used $T_c = 2.3$~K as measured for the sample studied in Ref.~\cite{Bhattacharyya2019}. That required taking slightly larger value of $\mu^* = 0.29$.}, and  
the BCS theory. 
 As in the case of the specific heat, the Eliashberg theory leads to a non-BCS curve, and the differences are more pronounced than in the case of $C_e$.
More importantly, the isotropic Eliashberg solution significantly deviates from the experimental data in both the lower and upper $T/T_c$ range. 
Much better agreement between the experimental data and theory was 
obtained in Ref.~\cite{Bhattacharyya2019} under the assumption of the $d$-wave gap symmetry.
{Thus, temperature dependence of both the electronic specific heat and magnetic field penetration depth cannot be described by the isotropic Eliashberg formalism
which supports the possibility of a non-$s$-wave gap symmetry in ThCoC$_2$.}

\begin{figure}[t]
\includegraphics[scale=.6]{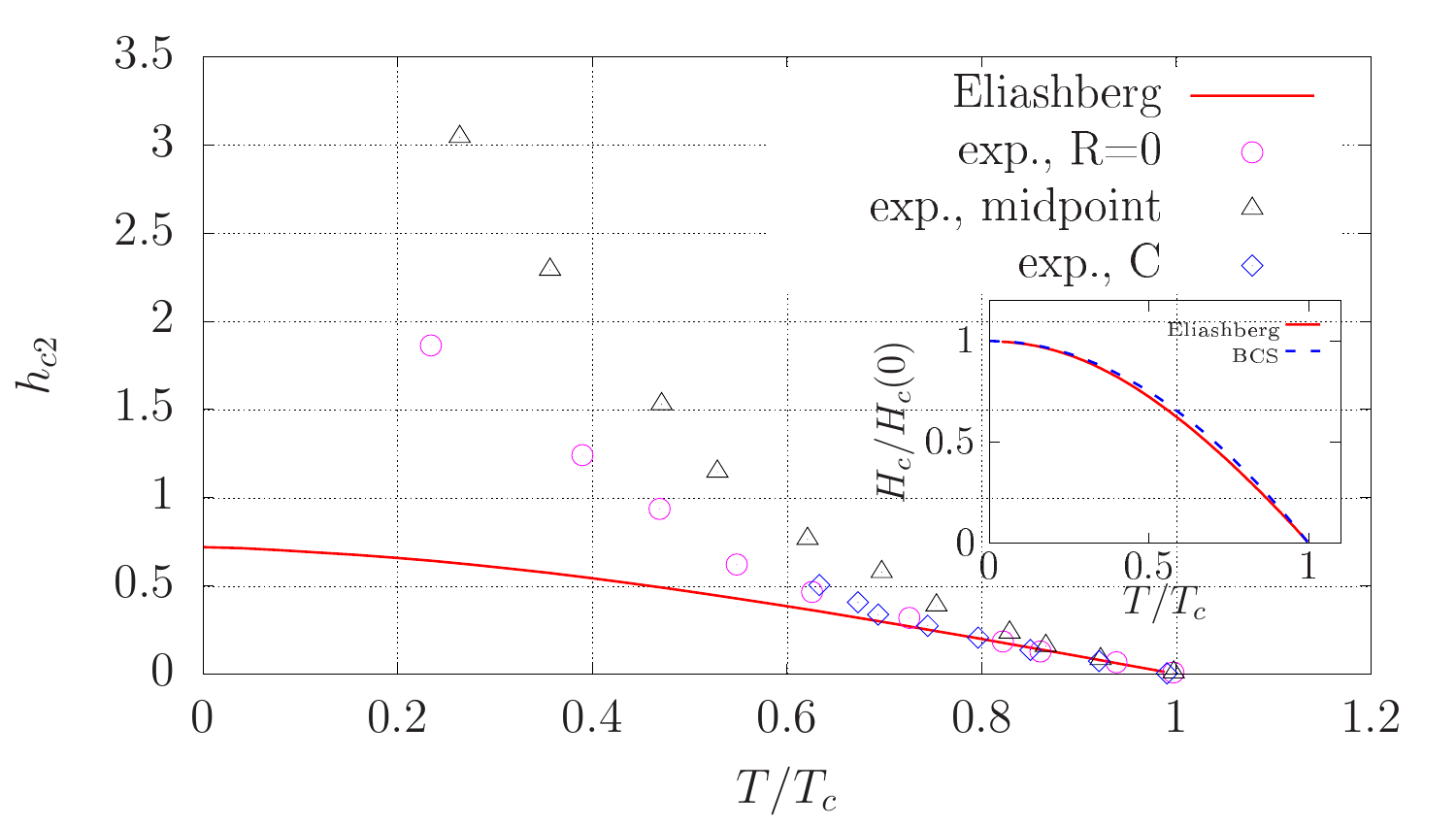}
\caption{  Upper critical field $h_{c2}=H_{c2}(T)/T_c H'_{c2}(T_c)$ as a function of the reduced temperature $T/T_c$, comparison with the experiment~[\onlinecite{Grant2014}]. Inset presents the comparison of the thermodynamic critical field calculated within the BCS and Eliashberg theories.} 
\label{fig:hc}
\end{figure}

A competitive concept on the nature of superconductivity in ThCoC$_2$ concerns multiband effects which are strongly indicted by the critical magnetic field measurements with the  change of the $H_{c2}(T)$ curvature reported in Ref.~[\onlinecite{Grant2014}].  
 The upper critical field within the Eliashberg model can be evaluated based on equations~\cite{Carbotte1990}
\begin{equation}
    \tilde{\Delta}(i\omega _n)=\pi k_BT \sum _{n'} \frac{\left [ \lambda(n-n')-\mu^* \theta(\omega_c - \omega _{n'}) \right ] \tilde{\Delta}(i\omega _{n'}) }{\chi ^{-1}[\tilde{\omega}(i\omega_{n'})]-\pi t^+},
\label{eq:delta_hc2}
\end{equation}
where
\begin{equation}
\tilde{\omega} (i\omega_{n})=\omega _n + \pi k_BT \sum _{n'}\lambda(n-n') \text{sgn}(\omega_{n'})+\pi t^+ \text{sgn}(\omega_{n'}),
\label{eq:omega_hc2}
\end{equation}
with $t^+=\frac{1}{2\pi \tau}$ where $\tau$ is the electronic scattering time due to the presence of impurities. 
In Eq.~(\ref{eq:delta_hc2}) the function $\chi(\tilde{\omega})$ is given by
\begin{equation}
    \chi (\tilde{\omega}(i\omega _n))=\frac{2}{\sqrt{\alpha}} \int _0 ^\infty dq \: e^{-q^2} \tan^{-1} \left ( \frac{\sqrt{\alpha}q}{|\tilde{\omega}(i\omega _n)|} \right ),
\end{equation}
with 
\begin{equation}
\alpha(T)=\frac{1}{2}|e|H_{c2}(T)v_F^2,   
\end{equation}
where $e$ is the elementary charge and $v_F$ is the Fermi velocity. 
To evaluate $\tau$
we have calculated the electrical conductivity $\sigma$ tensor for ThCoC$_2$ using the Boltzmann formalism in the constant scattering time approximation, as implemented in the {\sc boltztrap} code~\cite{boltztrap}. 
In this approach and assuming that the electronic scattering time $\tau$ is independent of {\bf k}-vector 
one may calculate the ratio of $\sigma/\tau$. The computed value, averaged over directions, is 
$\sigma/\tau = 17.3\times 10^{19}$ $\Omega^{-1}$ m$^{-1}$ s$^{-1}$. 
The measured residual resistivity for the polycrystalline sample of ThCoC$_2$ is $\rho_0 = 0.37$~$\mu\Omega$cm~\cite{Grant2014}, which gives relatively large value of $\tau = 1.56 \times 10^{-12}$ s which originates from the high quality of the sample studied in Ref.~\cite{Grant2014} with low amount of defects. 
Estimated value of $\tau$ results in $t^+ = 6.71\times 10^{-5}$ eV to be used in the above-mentioned formulas, and corresponds to the clean limit. 
The obtained temperature dependence of the normalized upper critical field $h_{c2}=H_{c2}(T) / T_c \frac{dH_{c2}(T_c)}{dT}$ is shown in Fig.~\ref{fig:hc} where the inset additionally presents the thermodynamic critical field $\frac{H_c}{\sqrt{N(E_F)}}=\sqrt{-8\pi\frac{\Delta F}{N(E_F)}}$ from the BCS model and Eliashberg theory. For comparison, the experimental data from 
different experimental methods of extraction, that is, from resistivity and specific heat measurements, are also presented. 

As expected, within the one-band Eliashberg model, the curvature of $h_{c2}$ is preserved in whole range of $T/T_c$ indicating  that the change of the curvature in the experimental curve may indeed result from multiband effects, not captured in the model presented here. Note moreover that the fully angle-resolved critical field measurements could shed light on the symmetry of the gap \cite{Wang2016} but they are yet to be done.

\subsection{Isotope effect}

\begin{figure}[t]
    \centering
    \includegraphics[width=0.90\columnwidth]{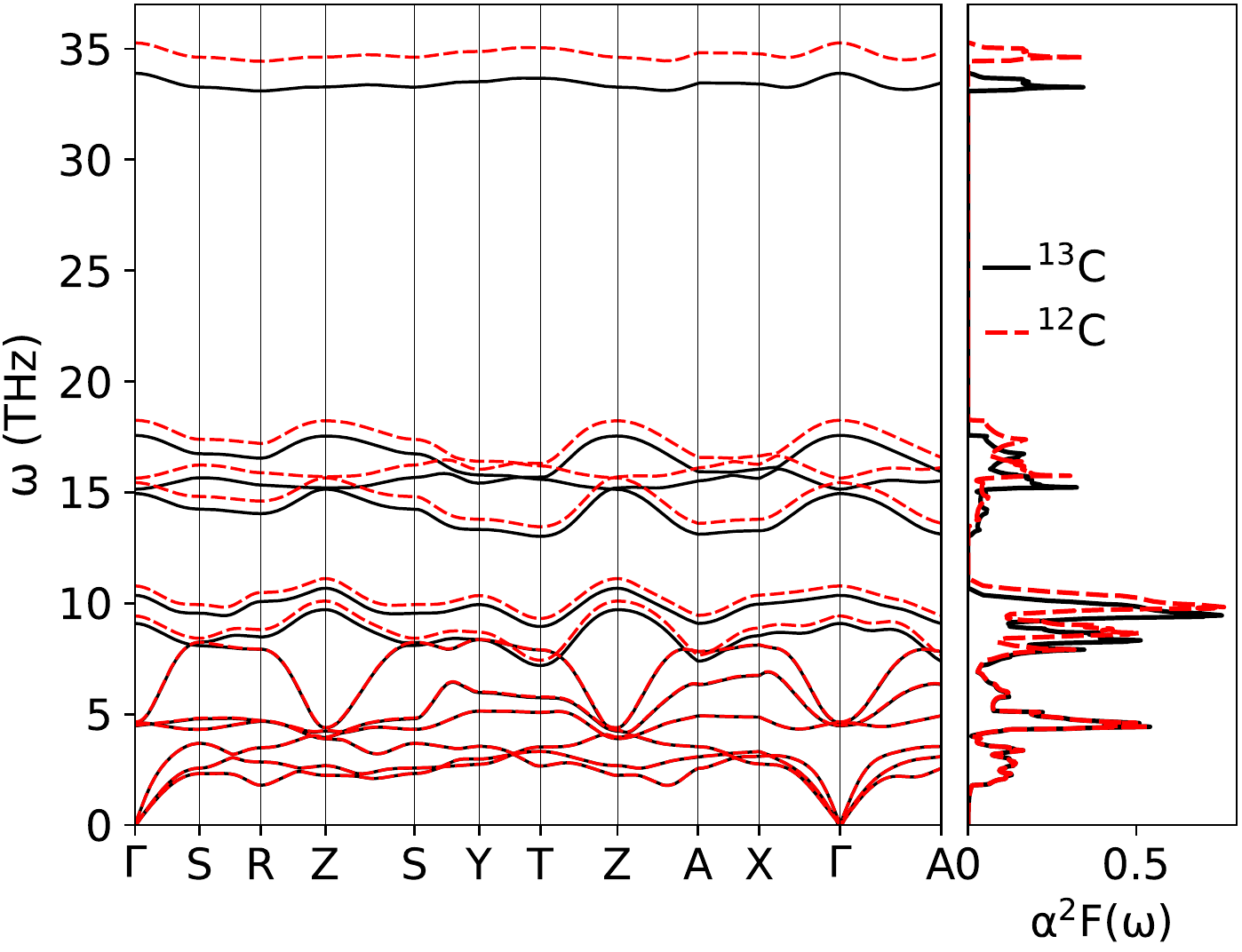}
    \caption{Phonon dispersion relations and Eliashberg functions of $\mathrm{ThCoC_2}$ calculated with $\mathrm{^{12}C}$ and $\mathrm{^{13}C}$ carbon isotopes.}
    \label{fig_a2F_C12_C13}
\end{figure}

The experiment which should help to distinguish between the electron-phonon pairing mechanism in ThCoC$_2$, supported by our calculations, {\it versus} spin fluctuation mechanisms proposed in Ref.~\cite{Bhattacharyya2019} is the measurement of the isotope effect. 
The isotope effect is usually described by the power law $T_c \propto M^{-\alpha}$, where M is the isotope's mass, and its presence clearly indicates the role of atomic vibrations in the superconductivity of the studied system. In our case of ThCoC$_2$, isotope effect can be measured only for carbon, which has two stable isotopes, $^{12}$C with the abundance of about 98.9\% and $^{13}$C with 1.1\%~\cite{carbon}.
The isotope effect exponent, however, may differ significantly from the model $\alpha = 0.50$ value, especially in the multi-atomic compounds, depending on the details of the phonon spectrum and electron-phonon interaction. For example, in $\mathrm{MgCNi_3}$~\cite{Klimczuk2004} a strong isotope effect was observed, as $T_c$ dropped by 0.3~K when $\mathrm{^{12}C}$ was fully replaced with $\mathrm{^{13}C}$, giving $\alpha=0.54(3)$. On the other hand, in borocarbide YNi$_2$B$_2$C, significant isotope effect was observed only for boron atoms, with $\alpha \simeq 0.2$, being negligible for carbon~\cite{borocarbides,borocarbides2}.
The difference in the atomic mass of the constituent elements is an important factor in determining the magnitude of the isotope effect in multiatomic compounds. This can be illustrated using the binary LiBi as an example, as it is built from the heaviest and the lightest stable solid elements. In this superconductor, the replacement of the natural Li (average mass 6.94 u) with the lighter $\mathrm{^{6}Li}$ increased $T_c$ by 0.02 K from 2.45 K to 2.47 K, resulting in an exponent $\alpha=0.04$~\cite{Takashi1971}. 
Even though the effect was very weak, it was possible to determine it experimentally {and theoretically~\cite{libi-gornicka}.
Although for the present case of ThCoC$_2$ due to its deviations form the isotropic Eliashberg superconductivity we cannot provide an accurate prediction on the change in $T_c$, we may simulate how the phonon spectra and electron-phonon coupling is modified upon isotope substitution and estimate the variation in $T_c$ using the conventional approach.}

\begin{figure}[t]
    \centering
    \includegraphics[width=0.90\columnwidth]{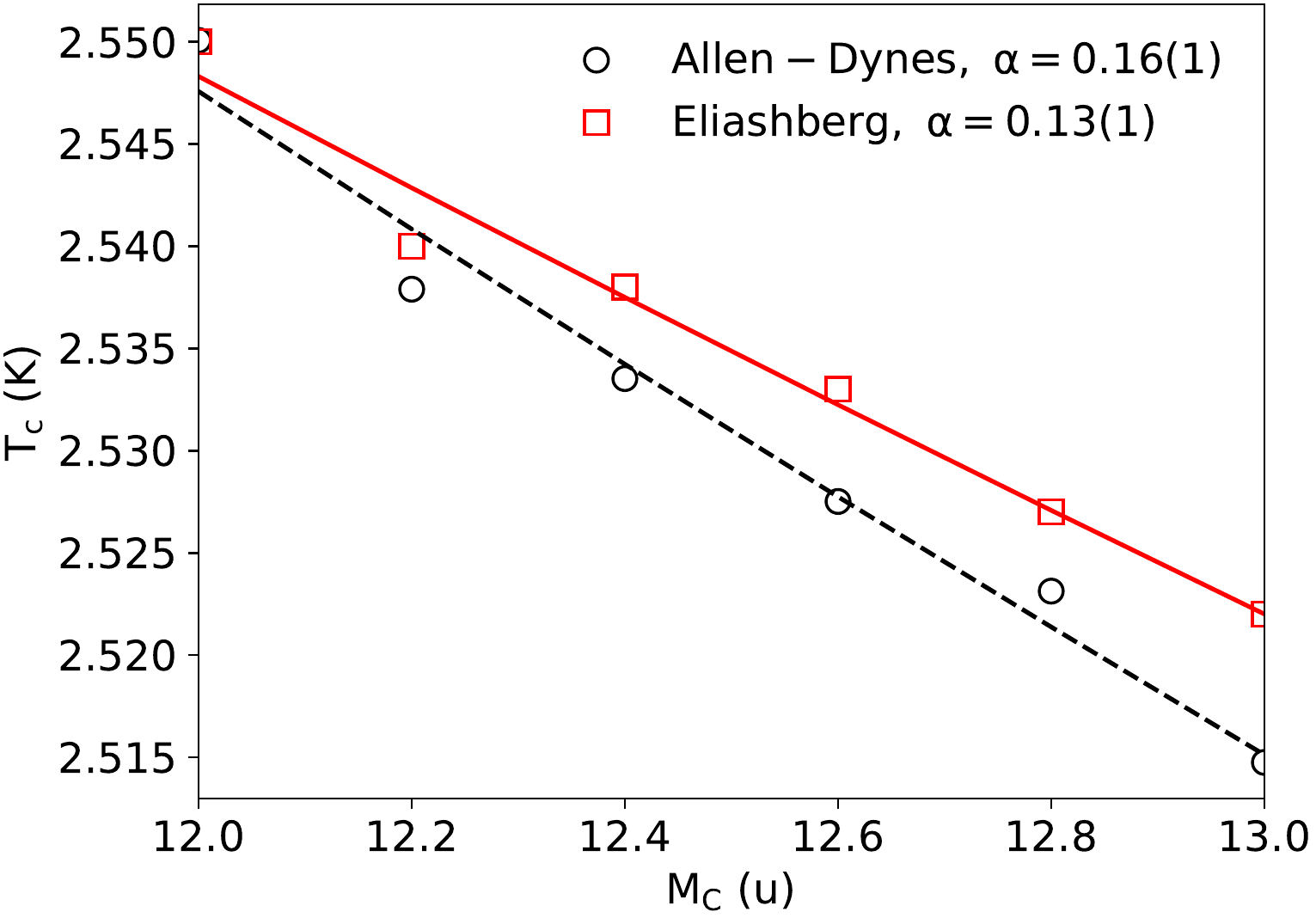}
    \caption{Critical temperature of $\mathrm{ThCoC_2}$ as a function of C atom's mass (open symbols) calculated using the Allen-Dynes formula or Eliashberg equations. Curves are the power law fits, $T_c = {\rm const}\times M^{-\alpha}$.}
    \label{fig_tc_vs_cmass}
\end{figure}

Phonon dispersion relations and Eliashberg functions calculated for $\mathrm{^{12}C}$ and $\mathrm{^{13}C}$ isotopes are shown in Fig.~\ref{fig_a2F_C12_C13}. 
Heavier C atoms resulted in the lowering of frequencies of the six highest optical modes of carbon, not affecting Co and Th modes. 
In addition, to simulate the measurements for the samples with a partial $\mathrm{^{12}C}$-$\mathrm{^{13}C}$ substitution, calculations were done with a C mass between 12 u and 13 u with a step of $0.2$~u, not shown here for the sake of clarity. 
With these new Eliashberg functions, the effect of the carbon mass on $T_c$ was simulated using both the Allen-Dynes formula and the Eliashberg formalism.
Other relevant parameters were unchanged and correspond to the values which are used to reproduce the experimental $T_c = 2.55$ K for the $\mathrm{^{12}C}$ case.
In the Eliashberg formalism, the cutoff frequency $\omega_c = 4\omega_{\rm max}$ was adopted according to the shift in the frequency of the highest carbon mode. 
Results are presented in Fig.~\ref{fig_tc_vs_cmass} and decrease of $T_c$ by  0.035 K for the full $\mathrm{^{13}C}$ substitution is noted.
The isotope effect exponents of $\alpha=0.13(1)$ and $\alpha=0.16(1)$ are predicted based on the Eliashberg formalism and Allen-Dynes formula, respectively.
When the Allen-Dynes formula is used, the drop in $T_c$ is intuitively explained as resulting from
the decrease in $\langle\omega_{\rm log}^{\alpha^2F}\rangle$ due to $\mathrm{^{13}C}$ substitution (5.816~THz to 5.717~THz) whereas $\lambda$ remained almost unchanged (increase from 0.5902 to 0.5906).
The predicted effect is not strong, but still larger than e.g. in LiBi, thus its experimental detection should be possible, once a set of samples with different $\mathrm{^{13}C}$ content is synthesized under the same conditions.

Two comments have to be added to the discussion of the isotope effect in ThCoC$_2$. 
First, the predicted shift in the phonon frequencies upon $\mathrm{^{13}C}$ substitution is an effect independent of superconductivity or gap symmetry.
Thus, regardless of the method used to determine $T_c$, if the electron-phonon coupling is the pairing interaction, the magnitude of the shift in $T_c$ should be reasonably predicted, since generally $T_c \propto \omega_c$ with $\omega_c$ being the characteristic phonon frequency.
Second, the sole observation of the isotope effect does not exclude the spin fluctuations as the pairing interaction, as the interplay between electron-phonon and spin-fluctuations leads to the presence of the isotope effect in cuprates, like YBCO~\cite{Nunner99}. 
There, the antiferromagnetic spin fluctuations are considered as the dominant pairing interaction, which compete with the weaker electron-phonon coupling, responsible for the isotope effect.
Spin fluctuations are predicted~\cite{Nunner99} to reduce the isotope effect exponent $\alpha$, with the resulting magnitude depending on the relative strength of spin fluctuations and electron-phonon coupling. 
In the case of $\mathrm{ThCoC_2}$, under the assumption of a dominant electron-phonon coupling and possible accompanying weak paramagnetic spin fluctuations, only a small renormalization of the isotope effect exponent from that determined by phonons can be expected~\cite{isotope_nonph}.
Concluding, an experimental analysis of the isotope effect should deliver important results for the determination of the pairing mechanism.

\section{Summary and discussion}
\label{sec:summary}

In summary, we have presented theoretical studies of the electronic structure, lattice dynamics, electron-phonon interaction, and superconductivity in noncentrosymmetric ThCoC$_2$. Calculations show that all atoms contribute to the Fermi surface, which in the scalar-relativistic case consists of two parts: a complex large hole-like sheet and a very small electron pocket.
Spin-orbit interaction splits the electronic bands with a quite large average energy splitting of $\overline{\Delta E_{SOC}} = 150$ meV, which leads to the appearance of two dominating hole-like Fermi surface sheets.
Due to the large differences in the atomic masses of the constituent atoms, in the phonon spectrum of ThCoC$_2$  we may distinguish the regions dominated by each of the atoms' vibrations: Th below 5 THz, Co between 5 and 8 THZ, and C between 8 and 35 THz.
The strongest electron-phonon interaction, in the sense of the largest phonon linewidths $\gamma_{\pmb{q}\nu}$ is associated with the carbon atoms' phonon modes. However, due to the lower phonon frequencies, the cobalt and thorium phonon modes also provide important contributions to the overall electron-phonon coupling parameter, calculated to be $\lambda = 0.59$. 
 As far as the phonon properties are concerned, a negligible effect of SOC was found both for phonon dispersion relations and electron-phonon coupling function $\alpha^2F(\omega)$.
Combination of the calculated density of states at the Fermi level, $N(E_F)$, and the electron-phonon coupling parameter $\lambda$ lead to the theoretical value of the renormalized Sommerfeld electronic specific heat coefficient $\gamma = \gamma_{\rm band}(1+\lambda) = 8.0$ (mJ mol$^{-1}$ K$^{-2}$), slightly lower than the experimental one, $\gamma_{\rm expt} = 8.4$ (mJ mol$^{-1}$ K$^{-2}$).
On the other hand, the superconducting critical temperature is overestimated, when calculated using the Allen-Dynes formula with the standard Coulomb pseudopotential parameter value (4.2 K for $\mu^* = 0.13$), and the experimental $T_c = 2.55$~K is reproduced for a relatively large value of $\mu^* = 0.165$. 
 
That may suggest the presence of other depairing mechanisms like electron-paramagnon interactions, which effectively enhance $\mu^*$. However, in connection with the unusual thermodynamic properties, it rather indicates that the superconductivity is not isotropic $s$-wave-like, and due to the strong spin-orbit coupling the Allen-Dynes formula may overestimate $T_c$.

{Additionally, the  thermodynamic  properties  of  the  superconducting phase in $\mathrm{ThCoC_2}$ were analyzed with the help of the isotropic Eliashberg equations.
A non-BCS temperature dependence of the electronic specific heat and the London penetration depth  were found. 
In contrary to the sister compound, $\mathrm{LaNiC_2}$, for which the Allen-Dynes formula worked well and  temperature dependence of the London penetration depth was close to that predicted by the isotropic Eliashberg formalism~\cite{lanic2}, in the case of $\mathrm{ThCoC_2}$ the experimental results are far from those predicted by the isotropic theory.
Similarly, the measured temperature dependence of the upper magnetic critical field is not explained in terms of the $s$-wave single-band model.
Those features are likely to be driven by the strong SOC found in ThCoC$_2$,
as in the limit of the strong spin-orbit interaction~\cite{mixedstate2}, the pairing inside the spin-split bands requires the odd parity of the gap with respect to the $\mathbf{k}\rightarrow -\mathbf{k}$.
This opens the possibility of a non-$s$-like gap symmetry in ThCoC$_2$, which is achievable even with the electron-phonon-based pairing~\cite{Karol-prb,Karol-physc,d-wave-elph}.
}

{
Concluding, in view of our results, the electron-phonon interaction is strong enough to mediate the superconductivity in ThCoC$_2$, but the measured thermodynamic properties of the superconducting phase much deviate from those predicted by the isotropic Eliashberg theory.
Further works are thus required to distinguish between the phonon and spin-fluctuation mechanisms, proposed along with the $d$-wave gap symmetry in Ref.~\cite{Bhattacharyya2019}.
Our calculations show that the analysis of the carbon isotope effect could yield important results. 
In spite of the large mass differences between Th, Co and C, observation of the isotope effect on C is predicted, with an exponent $\alpha \simeq 0.15$.
}

\section*{Acknowledgements}
This work was supported by the National Science Centre (Poland), project No. 2017/26/E/ST3/00119
and in part by the PL-Grid Infrastructure (allocation of computing time).
\bibliography{refs}

\end{document}